\documentclass[11pt,a4paper]{article}
\usepackage{jheppub}
\usepackage{latexsym}
\usepackage{pdflscape}
\usepackage{bm}
\numberwithin{equation}{section}
\allowdisplaybreaks

\allowdisplaybreaks[2]

\def\cO{\mathcal{O}}
\def\cJ{\mathcal{J}}
\def\cD{\mathcal{D}}
\def\cN{\mathcal{N}}
\def\cG{\mathcal{G}}

\def\mint{\int_{-\infty}^\infty\!\cdots\!\int_{-\infty}^\infty}
\def\l{\ell}

\newcommand{\be}{\begin{equation}}
\newcommand{\ee}{\end{equation}}
\newcommand{\ba}{\begin{aligned}}
\newcommand{\ea}{\end{aligned}}

\def\Res{\mathop {\rm Res} \limits}

\def\bra#1{\langle #1 |}
\def\ket#1{| #1 \rangle}

\DeclareMathOperator{\B}{B}

\def\({\left(}
\def\){\right)}

\newcommand{\pd}{\partial}
\DeclareMathOperator{\real}{Re}

\DeclareMathOperator{\Tr}{Tr}

\newcommand{\re}{{\rm e}}
\newcommand{\ri}{{\rm i}}
\newcommand{\rd}{{\rm d}}

\preprint{DESY\ 15-042}

\title{Spectral zeta function and non-perturbative effects in ABJM Fermi-gas}

\author{Yasuyuki Hatsuda}

\affiliation{DESY Theory Group, DESY Hamburg, \\
Notkestrasse 85, D-22603 Hamburg, Germany}

\emailAdd{yasuyuki.hatsuda@desy.de}

\abstract{
The exact partition function in ABJM theory on three-sphere can be regarded as a canonical partition function
of a non-interacting Fermi-gas with an unconventional Hamiltonian.
All the information on the partition function is encoded in the discrete spectrum of this Hamiltonian.
We explain how (quantum mechanical) non-perturbative corrections in the Fermi-gas system
appear from a spectral consideration.
Basic tools in our analysis are a Mellin-Barnes type integral representation and a spectral zeta function.
From a consistency with known results, we conjecture that the spectral zeta function in the ABJM Fermi-gas 
has an infinite number of ``non-perturbative'' poles,
which are invisible in the semi-classical expansion of the Planck constant.
We observe that these poles indeed appear after summing up perturbative corrections.
As a consequence, the perturbative resummation of the spectral zeta function causes 
non-perturbative corrections to the grand canonical partition function.
We also present another example associated with a spectral problem in topological string theory.
A conjectured non-perturbative free energy on the resolved conifold is successfully reproduced in this framework.
}

\begin{document}

\maketitle

\renewcommand{\thefootnote}{\arabic{footnote}}
\setcounter{footnote}{0}
\setcounter{section}{0}

\section{Introduction}\label{sec:intro}
In this work, we study non-perturbative aspects of the partition function in ABJM theory \cite{ABJM}, a 3d 
$\cN=6$ superconformal Chern-Simons-matter theory with $U(N)_k \times U(N)_{-k}$ gauge group.
The ABJM theory has many important features.
It is a low-energy effective theory on $N$ M2-branes put on a $\mathbb{C}^4/\mathbb{Z}_k$ singularity.
As a consequence, it has a gravity dual.
In the 't Hooft limit: $N \to \infty$ with $\lambda=N/k$ held finite, the dual description is type IIA string theory
on $AdS_4 \times \mathbb{CP}^3$.
In the limit: $N \to \infty$ with $k$ held finite, the gravity dual is M-theory on $AdS_4 \times S^7/\mathbb{Z}_k$.
We thus refer to this limit as the M-theory limit here.
The ABJM theory is a good example to probe M-theory.

After the seminal work by Pestun \cite{Pestun}, a localization technique was applied to many supersymmetric gauge theories
in various dimensions. For a recent review, see \cite{Teschner:2014oja} and related articles therein. 
Path integrals of partition function and some BPS quantities finally reduce to finite dimensional matrix integrals. 
We emphasize that these reductions are exact, and no information is lost.
In \cite{KWY1, Jafferis, HHL1} (see also \cite{Hosomichi} for a review), 
the localization was used in a wide class of gauge theories on three-sphere.
An advantage of the 3d theories is that their matrix integrals are much simpler than those in higher dimensions.
Therefore it is tractable to understand the large $N$ behavior at a very quantitative level.

Our starting point here is the exact partition function of ABJM theory \cite{KWY1}:
\be
Z_\text{ABJM}(N,k)=\frac{1}{N!} \int \frac{\rd^N \mu\rd^N \nu}{(2\pi)^{2N}}
\frac{\prod_{i<j} ( 2\sinh \frac{\mu_i-\mu_j}{2} )^2 (2\sinh \frac{\nu_i-\nu_j}{2} )^2}
{\prod_{i,j} (2\cosh \frac{\mu_i-\nu_j}{2} )^2}
\exp \left[ \frac{\ri k}{4\pi} \sum_{i=1}^N (\mu_i^2-\nu_i^2) \right],
\label{eq:Z-ABJM}
\ee
where we follow a convention in \cite{DMP1}.
A fundamental problem is to understand the large $N$ behavior of the free energy $\log Z_\text{ABJM}(N,k)$.
This still remains non-trivial after the localization.
In the 't Hooft limit, this was done in \cite{MP1, DMP1} in great detail.
After overcoming several technical issues, a recursion relation to determine the all-genus free energy was found \cite{DMP1}.
An important consequence of \cite{DMP1} is that the genus expansion of the free energy is very likely
Borel summable.
Naively, one might expect that the all-genus Borel resummation reproduces the exact partition function,
but this is not the case.
As observed in \cite{GMZ}, the Borel resummation does not agree with the exact answer.
The disagreement is caused by so-called complex instantons \cite{DMP2}.
The physical origin of such instantons in string theory is nothing but D-branes \cite{DMP2}. 
In summary, the ABJM partition function is Borel summable with respect to string coupling $g_s=2\pi/k$,
but  receives the non-perturbative corrections of form $\re^{-1/g_s}$.
The complete large $N$ expansion is not the genus expansion in $g_s$ but rather its trans-series expansion.
We note that there is a framework to explore non-perturbative effects systematically,
known as resurgence theory \cite{Ecalle}.
In matrix models and related topological strings, the resurgence analysis started from a pioneering work \cite{Marino2008},
based on \cite{Marino2006, MSW} (see also \cite{GIKM}).
Recently, it was developed in \cite{ASV, SV, SESV, ARS} systematically.
The resurgence should be useful to understand the complex instanton effects
in the ABJM matrix model.

Here we are rather interested in the M-theory limit.
In an earlier important work \cite{HKPT}, it was shown that the ABJM matrix model \eqref{eq:Z-ABJM}
behaves as $\log Z_\text{ABJM} \sim N^{3/2}$ in the large $N$ limit with fixed $k$.
This is perfectly consistent with the M-theory expectation.
However, the computation of the higher order $1/N$ corrections seems to be hard in this approach.

A resolution to overcome this difficulty is now known.
In \cite{MP2}, it was shown that the ABJM partition function \eqref{eq:Z-ABJM} is exactly equivalent to
a canonical partition function of an ideal Fermi-gas.
The Fermi-gas approach is very powerful in the M-theory limit,
and revealed a detailed non-perturbative structure in dual M-theory.
It turned out that the existence of two types of instantons, 
i.e., worldsheet instantons and membrane instantons, 
is crucially important for the non-perturbative definition of the theory.
In particular, the worldsheet instanton correction diverges at every physical value of
the coupling, and that divergence is precisely canceled by the
similar,  but with opposite sign, divergence of the membrane instanton correction \cite{HMO2}.
This pole cancellation mechanism is now observed in a wide class of 3d Chern-Simons gauge theories
\cite{HO, MN1, MN2, MN3} and also in topological strings \cite{HMMO, KM, GHM1}. 
It is conceptually important since
it guarantees a smooth interpolation between the weak coupling (type IIA) regime and the strong coupling
(M-theory) regime.
The Fermi-gas formulation is also useful in ABJ theory \cite{ABJ}, %a generalization of ABJM theory, 
studied in \cite{MaMo, HoO} (see also \cite{AHS, Honda}), 
in the topological string analysis \cite{GHM1, Kashaev:2015kha, MZ}
and in 3d mirror symmetries \cite{Drukker:2015awa}.

\vspace{3mm}

\paragraph{A brief sketch.}

In order to make a problem clearer, let us here sketch a basic story in the Fermi-gas approach.
A bit more detailed discussion will be reviewed in the next section.
Throughout this paper, we focus on a generating function $\Xi(\kappa,k)$ of \eqref{eq:Z-ABJM} 
rather than the partition function itself:
\be
\Xi (\kappa, k):=1+\sum_{N=1}^\infty \kappa^N Z_\text{ABJM}(N,k).
\label{eq:Xi}
\ee
In the Fermi-gas picture, this is just the grand canonical partition function with fugacity $\kappa$.
A crucial consequence of \cite{MP2} is that this grand partition function is given by
a Fredholm determinant for density operator $\hat{\rho}$ describing the system.
As explained in \cite{MP2}, the density operator $\hat{\rho}$ is of trace class, and has an infinite number of discrete eigenvalues.
It is well-known that Fredholm determinants are entire functions and have an infinite number of zeros.
These zeros just corresponds to the eigenvalues of $\hat{\rho}$.
Importantly, in the Fermi-gas picture, the Chern-Simons level $k$ plays the role
of the Planck constant.
Therefore the semi-classical limit corresponds to $k \to 0$, which is the \textit{strong coupling} limit in ABJM theory.
Thanks to this remarkable correspondence, the semi-classical analysis tells us the information on the membrane instantons.
It is easy to see that the grand potential (the logarithm of the Fredholm determinant) is written as
\be
\cJ(\kappa,k):=\log \Xi (\kappa,k)=-\sum_{\ell=1}^\infty \frac{(-\kappa)^\ell}{\ell} Z(\ell),
\label{eq:J}
\ee
where $Z(s)$ is a \textit{spectral zeta function}, defined by
\be
Z(s):=\sum_{n=0}^\infty \lambda_n^s.
\label{eq:spec-zeta}
\ee
In the following analysis, this function plays a crucial role.
Here $\lambda_n$ ($n=0,1,2,\dots$) are the eigenvalues of $\hat{\rho}$ (see \eqref{eq:eigen-eq}).
Note that the spectral zeta function $Z(s)$ depends on $k$.
Also, note that the sum \eqref{eq:spec-zeta} converges only for $\real s>0$, but it is analytically continued to the whole complex plane,
as in the Riemann zeta function.
The relation \eqref{eq:J} is already useful to compute the partition function for small $N$.
In fact, using this relation, the exact values of $Z_\text{ABJM}(N,k)$ were computed for various $N$ and $k$ 
\cite{HMO1, PY, HMO2}. 
To obtain the large $N$ result, we need to go to the large $\kappa$ regime.
In this regime, the grand potential takes the form \cite{MP2}
 \be
\cJ(\mu,k)=\frac{2\mu^3}{3\pi^2 k}+\(\frac{1}{3k}+\frac{k}{24}\)\mu+A_\text{c}(k)+\cJ_\text{np}(\mu,k),\qquad
\kappa=\re^{\mu},
\label{eq:J-large}
\ee
where we have introduced a chemical potential $\mu$ for later convenience.
In the following, we will use both $\kappa$ and $\mu$, interchangeably.
From this result, one can immediately deduce the $N^{3/2}$ behavior of the free energy at large $N$ 
and also the Airy function behavior first
derived in \cite{FHM}.
The constant part $A_\text{c}(k)$ is a complicated function of $k$, but its exact form is known \cite{KEK}
(see \eqref{eq:A-int}).
The last part $\cJ_\text{np}(\mu,k)$ is exponentially suppressed in $\mu \to \infty$.\footnote{%
The subscript ``np'' means the non-perturbative corrections in $\mu$.
In most of this paper, the term ``non-perturbative'' means the non-perturbative corrections in $k$.
In this terminology, the correction for $m=0$ in \eqref{eq:J-np} is indeed perturbative.
}
As studied in \cite{HMO2, CM, HMO3}, this part turned out to be
\be
\cJ_\text{np}(\mu,k)=\sum_{(\l,m) \ne (0,0)}
f_{\ell,m}(\mu,k) \exp \left[ -\( 2\ell+\frac{4m}{k} \)\mu \right],
\label{eq:J-np}
\ee
where all the coefficients $f_{\ell,m}(\mu,k)$ can be computed with the help of the refined topological string
on local $\mathbb{P}^1 \times \mathbb{P}^1$, in principle \cite{HMMO}.\footnote{%
Strictly speaking, the grand potential receives an additional contribution to the topological string prediction.
This was first observed in \cite{HMO2}, and is now understood as a contribution from a generalization
of Jacobi theta function \cite{CGM, GHM1, GHM2}. Here we do not care about it because it is not important in the following analysis.
The important fact is that this contribution does not change the form \eqref{eq:J-np}.}
Physically, the correction $\cO(\re^{-2\mu})$ corresponds to the membrane instanton correction,
while the correction $\cO(\re^{-4\mu/k})$ to the worldsheet instanton correction.
Note that the corrections for $m \ne 0$ are invisible in the semi-classical expansion around $k=0$ due to the exponentially
suppressed correction $\cO(\re^{-4\mu/k})$.
These are understood as \textit{(quantum mechanical) non-perturbative corrections in $k$}.

The important point is that it is far from obvious to go from the small $\kappa$ result \eqref{eq:J}
to the large $\kappa$ result \eqref{eq:J-large} and vice versa.
One needs an analytic continuation to connect these two regimes.
In the semi-classical limit $k \to 0$, it was done in \cite{MP2}.
Also, at some special values of $k$ ($k=1,2,4$), 
the large $\mu$ expansion \eqref{eq:J-large} drastically simplifies,
and we can write it down in a closed form \cite{CGM, GHM2}.
In these special cases, it is possible to analytically continue the large $\kappa$ result to the small $\kappa$ regime,
and it reproduces all the known results.
However this has not been done for general $k$ so far.
This problem is important because it is equivalent to interpolate between the small $N$ result and the large $N$ result,
including all the non-perturbative corrections, directly.
See also \cite{CSSV} for the related problem but in different setups and approach.

\begin{figure}[tb]
\begin{center}
\begin{tabular}{cc}
%\hspace{-3mm}
\resizebox{45mm}{!}{\includegraphics{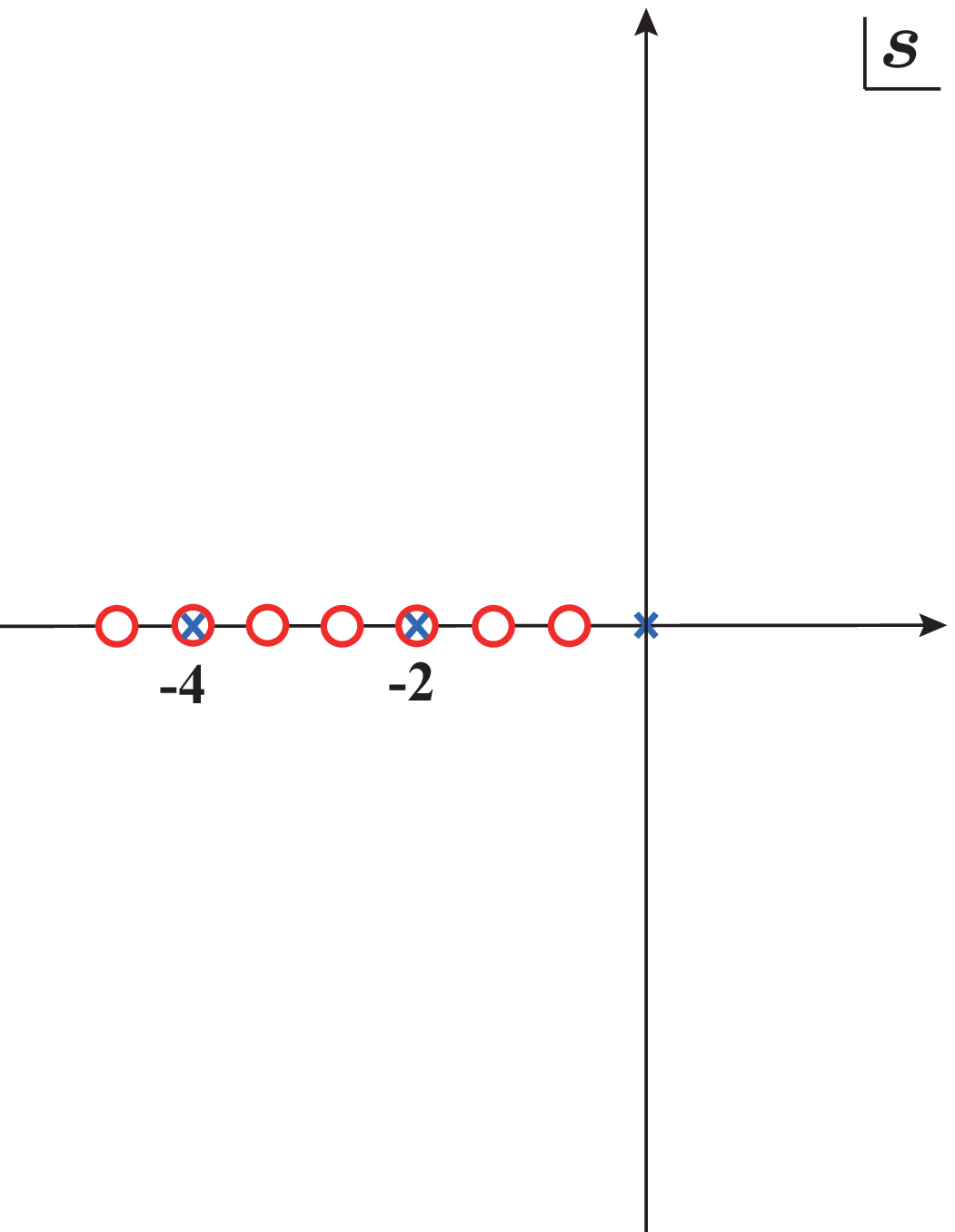}}
\hspace{10mm}
&
\resizebox{65mm}{!}{\includegraphics{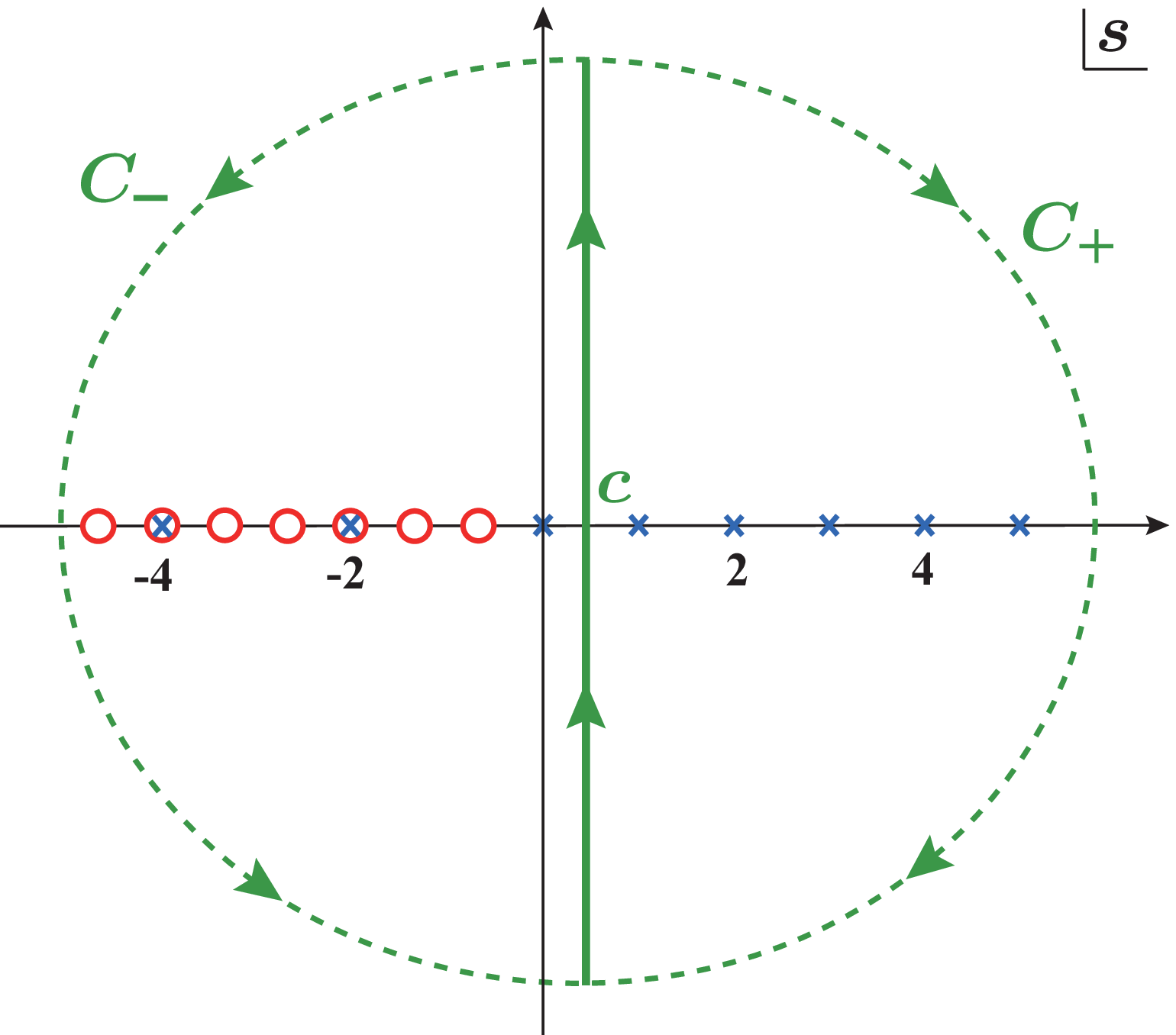}}
\vspace{-5mm}
\end{tabular}
\end{center}
  \caption{(Left) (Expected) pole structure of $Z(s)$ at $k=6$ is sketched. The blue ``$\times$'' show  ``perturbative'' poles while
the red  ``$\circ$'' show ``non-perturbative'' poles. In this case, the non-perturbative poles appear at $s=-2/3,-4/3,-2,\dots$.
The non-perturbative poles at $s=-2,-4,\dots$ are overlapped with the perturbative poles.
At these point, the pole cancellation \cite{HMO2} in the grand potential occurs.
(Right) Contour deformations of the Mellin-Barnes integral \eqref{eq:J-MB}. For $\mu<0$ ($\mu>0$), 
we can add the semi-circle $C_+$ ($C_-$). We also show the pole structure of the integrand in \eqref{eq:J-MB} at $k=6$. 
To pick up the poles correctly, we have to choose $c$ such that $0<c<1$.}
  \label{fig:Poles}
\end{figure}

In this paper, we give a clue to resolve this problem.
Our basic tool is a \textit{Mellin-Barnes type integral representation}.\footnote{%
We thank Marcos Mari\~no for telling us about a powerfulness of the Mellin-Barnes representation.}
As in hypergeometric series, the infinite sum \eqref{eq:J} admits the following integral representation\footnote{%
This is essentially an inverse Laplace transform. For a relationship between spectral zeta functions and spectral determinants,
see \cite{Voros}, for instance. 
}:
\be
\cJ(\mu,k)=-\int_{c-\ri \infty}^{c+\ri \infty} \frac{\rd s}{2\pi \ri} \Gamma(s)\Gamma(-s)Z(s)\re^{s \mu},
\label{eq:J-MB}
\ee
where $c$ is a constant, which must be chosen in $0<c<1$ in the current case, as explained just below.
To go from \eqref{eq:J} to \eqref{eq:J-MB}, one must be careful about the integration contour and the pole structure of the integrand.
As discussed later, it turns out that the spectral zeta function does not have any poles in $\real s>0$.
In $\mu<0$, we can deform the contour of \eqref{eq:J-MB} by adding an infinite semi-circle $C_+$ as shown in figure~\ref{fig:Poles}.
Then the integral can be evaluated by the sum of the residues over all the poles in $\real s>c$.
If $0<c<1$, the poles in this regime are at $s=\ell$ ($\ell \in \mathbb{Z}_{>0}$), and thus we precisely recover the sum \eqref{eq:J}.
On the other hand, for $\mu>0$ we can deform the contour by adding the opposite semi-circle $C_-$ as in figure~\ref{fig:Poles}.
In this case, one needs the information of the poles in $\real s<c$.
In this regime, $Z(s)$ may have non-trivial poles,
and the problem is highly non-trivial.
%
%%%%%%%%%%%%%%%%%%%%%%%%
%%%%%%%%%%%%%%%%%%%%5
%
From the Mellin-Barnes integral point of view, the large $\mu$ expansion \eqref{eq:J-large} should be understood 
as the sum of the residues over all the poles in $\real s<c$.
More explicitly, the large $\mu$ expansion, from \eqref{eq:J-MB}, should be given by
\be
\cJ(\mu,k)=-\sum_{\substack{\text{All the poles}\\\text{in}\,\real s<c}}\Res_s \Gamma(s)\Gamma(-s)Z(s)\re^{s \mu}, \qquad (\mu \to \infty).
\ee
As we will see in the next section, the polynomial part in \eqref{eq:J-large} indeed comes from the residue at $s=0$.
The contributions from the poles in $\real s<0$ are exponentially suppressed in $\mu \to \infty$.
Therefore we conclude that 
\textit{to reproduce the result (\ref{eq:J-large}) with (\ref{eq:J-np}), the integrand of (\ref{eq:J-MB}) must have the poles (only) at}
\be
s=-2\ell-\frac{4m}{k},\qquad \ell,m=0,1,2,\dots.
\label{eq:poles}
\ee
This is a consistency requirement from the known results, and must be confirmed by a direct analysis.
Since the gamma function $\Gamma(\pm s)$ has poles only at $s=0,\mp 1, \mp 2,\dots$, respectively,
we conclude that most of the poles in \eqref{eq:poles} must come from the spectral zeta function $Z(s)$.
We note that the poles for $m \ne 0$ go to the infinity in the limit $k \to 0$, and they are invisible
in the semi-classical expansion. In this sense, we call these poles \textit{non-perturbative poles} here.%
\footnote{A concept of non-perturbative poles is also found 
in a similar but different context \cite{AHS, HNS}.
We note that here we consider the small $k$ expansion while there the large $k$ (or small $g_s$) expansion
was considered.
}
In the semi-classical expansion, only the poles with $m=0$ appear.
At present we have no way to determine the complete pole structure of $Z(s)$, but
in this paper we give an evidence that it has at least a non-perturbative pole at $s=-4/k$.
The Mellin-Barnes representation states that this pole produces the correction $\re^{-4\mu/k}$,
which is nothing but the leading non-perturbative correction to the grand potential.
This idea is widely applicable to other examples.
In fact, in addition to the ABJM Fermi-gas, we also pick up another example, in which this mechanism works.
This example is related to a spectral problem in topological string theory \cite{ACDKV, Huang:2014eha, GHM1}.
We show that a conjectured non-perturbative correction to the free energy \cite{HMMO} on the resolved conifold
is indeed reproduced in this approach.

\begin{figure}[tb]
\begin{center}
\includegraphics[height=6.5cm]{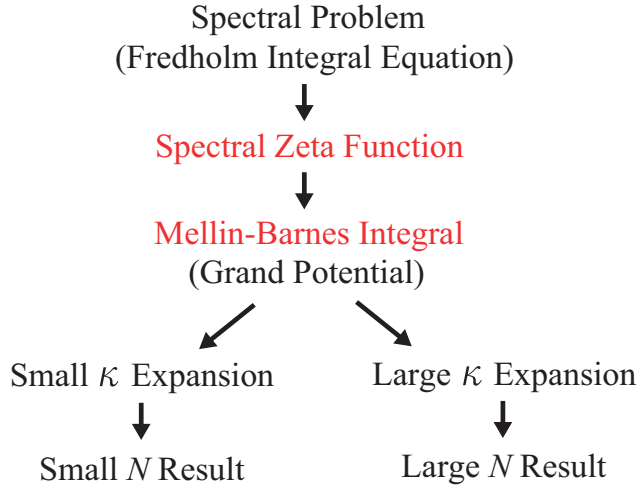}
\end{center}
  \caption{Basic flow of our analysis. This approach is widely applicable to other examples 
with a well-defined spectral problem.}
  \label{fig:strategy}
\end{figure}

Our basic strategy in this paper is schematically shown in figure~\ref{fig:strategy}.
The Mellin-Barnes type representation \eqref{eq:J-MB} is very involving. It is clear that all the information 
on the (grand) partition function
is encoded in the spectral zeta function in the whole regime of $\mu$ (or $N$). 
In particular, to extract the large $\mu$ result, it is important to understand the pole structure of $Z(s)$.
We observe that the resummation of the semi-classical expansion of $Z(s)$ produces the non-perturbative pole at $s=-4/k$.
This is conceptually interesting. We only consider the semi-classical analysis on the spectral zeta function.
Nevertheless we can explain the appearance of the non-perturbative correction to the grand potential.%
\footnote{%
There is, however, a possibility that the spectral zeta function itself receives non-perturbative corrections.
If so, these corrections might also have the poles \eqref{eq:poles}.
We  have no evidence of the existence of such corrections, but cannot exclude this possibility in this work.
We thank Marcos Mari\~no for pointing out this issue.
}
This is because that the Mellin-Barnes integration does not commute with the WKB infinite series.
\textit{The perturbative resummation of the spectral zeta function
causes the non-perturbative corrections to the grand potential.}
This is a main message of this paper.

\vspace{3mm}

\paragraph{Outline.}

The organization of this paper is as follows.
In section~\ref{sec:review}, we briefly review the Fermi-gas approach proposed in \cite{MP2}.
In particular, we see that the Mellin-Barnes representation \eqref{eq:J-MB} is indeed powerful to
know the grand potential in the large $\mu$ limit.
In section~\ref{sec:toy}, we propose a toy model, which has many common features to the ABJM Fermi-gas system.
Through this example, we can learn how the non-perturbative corrections to the grand potential appear
via the Mellin-Barnes representation.
In section~\ref{sec:ABJM-zeta}, we proceed to the analysis of the ABJM spectral zeta function.
First, we develop an efficient way to compute the semi-classical expansion systematically.
Then, using this result, we show that the spectral zeta function indeed has the non-perturbative pole at $s=-4/k$.
In section~\ref{sec:top}, we give another example, which is closely related to a spectral problem for the topological string
theory \cite{GHM1}. We see that our prescription here indeed reproduces the non-perturbative free energy
of the resolved conifold.
In section~\ref{sec:conclusion}, we give conclusions.
Appendix~\ref{sec:Wigner} is a detailed computation of the Wigner transform, which is used in section~\ref{sec:ABJM-zeta}.
In appendix~\ref{sec:diff}, we give a few results on the differential operators, which are useful to compute the quantum
corrections to the grand potential and to the spectral zeta function.

\section{Review of the ABJM Fermi-gas}\label{sec:review}
Let us quickly review the Fermi-gas formulation in \cite{MP2}.
We start with the exact partition function \eqref{eq:Z-ABJM}.
The crucial step is to rewrite this matrix integral as the following form
\be
Z_\text{ABJM}(N,k)=\frac{1}{N!} \sum_{\sigma \in S_N} (-1)^\sigma \int_{-\infty}^\infty \rd^N x\,
\prod_{i=1}^N \rho (x_i, x_{\sigma(i)}),
\label{eq:Z-FG}
\ee
where
\be
\rho(x_1,x_2)=\bra{x_1} \hat{\rho} \ket{x_2} =\frac{1}{2\pi k} \frac{1}{(2\cosh \frac{x_1}{2})^{1/2}}
\frac{1}{(2\cosh \frac{x_1-x_2}{2k}) } \frac{1}{(2\cosh \frac{x_2}{2})^{1/2}}.
\ee
Importantly, the expression \eqref{eq:Z-FG} can be regarded as the canonical partition function of an ideal Fermi-gas
described by the unconventional density operator $\hat{\rho}$.
The quantum Hamiltonian of this system is given by
\be
\re^{\hat{H}}=\hat{\rho}^{-1}=\(2\cosh \frac{\hat{x}}{2}\)^{1/2}\( 2\cosh \frac{\hat{p}}{2} \) \(2\cosh \frac{\hat{x}}{2}\)^{1/2},
\ee
where $\hat{x}$ and $\hat{p}$ are the canonical variables satisfying the commutation relation
\be
[\hat{x},\hat{p}]=\ri \hbar, \qquad \hbar=2\pi k.
\label{eq:comm-rel}
\ee
Therefore in this formulation, the Chern-Simons level $k$ plays the role of the Planck constant.
The semi-classical limit is $k \to 0$, which is a very strong coupling limit in ABJM theory.
Since the operator $\hat{\rho}$ is a non-negative, Hermitian and Hilbert-Schmidt operator,
it has positive discrete spectra $\lambda_0 \geq \lambda_1 \geq \lambda_2 \geq \cdots >0$. 
As already noted in the introduction, for the canonical partition function \eqref{eq:Z-FG},
the grand partition function \eqref{eq:Xi} is given by the Fredholm determinant:
\be
\Xi(\kappa,k)=\det(1+\kappa \hat{\rho})=\prod_{n=0}^\infty (1+\kappa \lambda_n)
=\prod_{n=0}^\infty (1+\re^{\mu-E_n}),\qquad \lambda_n=\re^{-E_n},
\label{eq:Xi-det}
\ee
where $E_n$ is the $n$-th energy eigenvalue of the system.
The eigenvalue problem is now given by
\be
\hat{\rho}\ket{\phi_n}=\lambda_n \ket{\phi_n}, \qquad (n=0,1,2,\dots).
\ee
As shown in \cite{MP2}, this can be recast in a homogeneous  Fredholm integral equation of the second kind in the coordinate representation:
\be
\int_{-\infty}^\infty \rd x' \, \rho(x, x') \phi_n(x')=\lambda_n \phi_n (x).
\label{eq:eigen-eq}
\ee
One can easily rewrite \eqref{eq:Xi-det} as the form in \eqref{eq:J}.
Therefore if one knows all the eigenvalues in \eqref{eq:eigen-eq}, one can compute the spectral zeta function, in principle.
However this is not easy for generic $k$.

The first step is to consider the semi-classical approximation, as in quantum mechanics.
Let us consider the semi-classical limit $k \to 0$.
At the leading approximation, we can treat $\hat{x}$ and $\hat{p}$ as commuting variables.
The classical Hamiltonian is thus given by
\be
\re^{H_\text{cl}}=\(2\cosh \frac{x}{2}\) \( 2\cosh \frac{p}{2}\).
\ee
Then the spectral zeta function for $s>0$ is easily computed by the phase space integral \cite{MP2}:
\be
Z^{(0)}(s)=\int_{-\infty}^\infty \frac{\rd x \rd p}{4\pi^2} \re^{-s H_\text{cl}}
=\frac{1}{4\pi^2} \frac{\Gamma^4(s/2)}{\Gamma^2(s)}.
\label{eq:Z0}
\ee
This is easily analytically continued to complex $s$.
Substituting this result into \eqref{eq:J}, one gets an analytic form of the grand potential in terms of
the hypergeometric functions \cite{MP2}.
Then, it is possible to analytically continue the grand potential to the large $\mu$ regime.
Here we use the Mellin-Barnes integral \eqref{eq:J-MB} to extract the large $\mu$ expansion directly from the spectral zeta
function.%
\footnote{This approach has already been discussed by Marcos Mari\~no in his review article \cite{MarinoReview}.}
The classical grand potential is represented by
\be
\cJ^{(0)}(\mu)=-\frac{1}{4\pi^2} \int_{c-\ri \infty}^{c+\ri \infty} \frac{\rd s}{2\pi \ri} \Gamma(-s) \frac{\Gamma^4(s/2)}{\Gamma(s)}
\re^{s\mu},
\label{eq:J0-MB}
\ee
where as explained in the previous section, we have to choose $c$ such that $0<c<1$.
It is obvious that the integrand in \eqref{eq:J0-MB} has poles at $s=-2n$ ($n=0,1,2,\dots$) in the $\real s<c$ region.
For the pole at $s=0$, one obtains the contribution
\be
-\frac{1}{4\pi^2} \Res_{s=0}\Gamma(-s) \frac{\Gamma^4(s/2)}{\Gamma(s)}
\re^{s\mu}=\frac{2\mu^3}{3\pi^2}+\frac{\mu}{3}+\frac{2\zeta(3)}{\pi^2}.
\ee
Similarly, from the poles at $s=-2n$ ($n \geq 1$), the exponentially suppressed corrections with $\re^{-2n \mu}$ 
are obtained.
We conclude that the classical grand potential has the following large $\mu$ expansion:
\be
\cJ^{(0)}(\mu)=\frac{2\mu^3}{3\pi^2}+\frac{\mu}{3}+\frac{2\zeta(3)}{\pi^2}+\( -\frac{4\mu^2}{\pi^2}+\frac{4\mu}{\pi^2}
+\frac{4}{\pi^2}-\frac{2}{3}\)\re^{-2\mu}+\cO(\mu^2 \re^{-4\mu}) \quad (\mu \to \infty).
\ee
This, of course, reproduces the result in \cite{MP2}.
We stress that to obtain this result, one does not have to perform the sum \eqref{eq:J}.
The Mellin-Barnes integral \eqref{eq:J-MB} allows us to know the grand potential in the large $\mu$ regime
directly from $Z(s)$.
As explained in section~\ref{sec:ABJM-zeta} in detail, it is possible to compute the semi-classical corrections
to the spectral zeta and the grand potential systematically.
These quantities admit the WKB expansions
\be
Z_\text{WKB}(s)=\frac{1}{k} \sum_{n=0}^\infty k^{2n} Z^{(n)}(s),\qquad
\cJ_\text{WKB}(\mu,k)=\frac{1}{k} \sum_{n=0}^\infty k^{2n} \cJ^{(n)}(\mu).
\label{eq:WKB}
\ee
As in \eqref{eq:Z-WKB}, the first correction $Z^{(1)}(s)$ is given by
\be
Z^{(1)}(s)=\frac{\pi^2}{96}\frac{s^2(1-s)}{ (1+s)} Z^{(0)}(s).
\ee
Using the Mellin-Barnes representation again, one easily finds
\be
\cJ^{(1)}(\mu)=\frac{\mu}{24}-\frac{1}{24}+\( \frac{\mu^2}{2}-\frac{5\mu}{6}-\frac{1}{12}+\frac{\pi^2}{12} \) \re^{-2\mu}
+\cO(\mu^2 \re^{-4\mu}),\qquad (\mu \to \infty).
\ee
Pushing the same computation in the higher corrections, we observe that these corrections
start from the constant term, i.e.
\be
\cJ^{(n)}(\mu)=\cO(1),\qquad n \geq 2, \qquad (\mu \to \infty).
\ee
Thus the WKB expansion of the grand potential is generically given by
\be
\cJ_\text{WKB}(\mu,k)=\frac{2\mu^3}{3\pi^2 k}+\(\frac{1}{3k}+\frac{k}{24}\)\mu+A_\text{c}(k)
+\sum_{\ell=1}^\infty (a_\ell(k)\mu^2+b_\ell(k)\mu+c_\ell(k))\re^{-2\ell \mu},
\label{eq:J-WKB-large}
\ee
where the constant term has the non-trivial corrections in all orders in small $k$ expansion.
It is not easy to determine its exact form only from the semi-classical analysis, 
but the all-order prediction is now known \cite{KEK} because this part corresponds to the constant map contribution
in the topological strings.
The all-order small $k$ expansion is given by
\be
A_\text{c}(k)=\frac{2\zeta(3)}{\pi^2 k}+\frac{1}{\pi^2 k}\sum_{n=1}^\infty \frac{(-1)^n}{(2n)!}B_{2n-2}B_{2n}(\pi k)^{2n},
\ee
where $B_n$ is the Bernoulli number.
Obviously this sum is a divergent series, but Borel summable for $k \in \mathbb{R}$.
After resumming it, one finally obtains the following integral representation \cite{HO}:
\be
A_\text{c}(k)=\frac{2\zeta(3)}{\pi^2 k}\(1-\frac{k^3}{16}\)
+\frac{k^2}{\pi^2} \int_0^\infty dx \frac{x}{e^{k x}-1}\log(1-e^{-2x}).
\label{eq:A-int}
\ee
In the above semi-classical analysis, only the correction with $m=0$ in \eqref{eq:J-np} appears.
In fact, at each order in the WKB expansion, we encounter only the poles at $s=-2n$ ($n=0,1,2,\dots$) for $\real s<c$.
Physically, these corrections are interpreted as the membrane instanton corrections \cite{MP2}.
We note that all of the coefficients $a_\ell(k)$, $b_\ell(k)$ and $c_\ell(k)$ in \eqref{eq:J-WKB-large} are exactly predicted from
the refined topological string on local $\mathbb{P}^1 \times \mathbb{P}^1$ \cite{HMMO}, in principle.
The other corrections in \eqref{eq:J-np} appear as non-perturbative corrections in $k$.
These are due to the worldsheet instanton corrections and bound states of the two kinds of instantons. 
The main purpose in this work is to explore a mechanism how these non-perturbative corrections are produced.

Remarkably, if introducing the following redefined chemical potential
\be
\mu_\text{eff}:=\mu+\frac{\pi^2 k}{2} \sum_{\ell=1}^\infty a_\ell(k) \re^{-2\ell \mu},
\label{eq:mu-eff}
\ee
then the grand potential \eqref{eq:J-large} drastically simplifies\footnote{%
As noted in the introduction, strictly, we need to subtract an ``oscillatory part'' in \cite{HMO2}.
Here we do not care about it. However, it will be important when one wants to compare the coefficients
at higher orders.}  
\cite{HMO3}
\be
\cJ(\mu,k)=\frac{2\mu_\text{eff}^3}{3\pi^2 k}+\(\frac{1}{3k}+\frac{k}{24}\)\mu_\text{eff}+A_\text{c}(k)+\cJ_\text{M2}(\mu_\text{eff},k)
+\cJ_\text{WS}(\mu_\text{eff},k),
\ee
where
\be
\ba
\cJ_\text{M2}(\mu,k)&= \sum_{\ell=1}^\infty (\widetilde{b}_\ell(k)\mu
+\widetilde{c}_\ell(k)) \re^{-2\ell \mu},\\
\cJ_\text{WS}(\mu,k)&=\sum_{m=1}^\infty d_m(k) \re^{-\frac{4m}{k}\mu}.
\ea
\label{eq:J-M2WS}
\ee
In the remaining sections, we see that the resummation of the perturbative WKB sum \eqref{eq:WKB} of $Z(s)$
causes the non-perturbative poles. 
These poles are sources of the non-perturbative corrections to the grand potential in \eqref{eq:J-np} or \eqref{eq:J-M2WS}.

\section{A toy model}\label{sec:toy}
In order to understand the essence,
let us start with a very simple toy model.
We consider the following \textit{exact} spectral zeta function:
\be
Z^\text{toy}(s)=\frac{1}{4\pi^2 s} \B\(\frac{k s}{4},\frac{ks}{4} \),
\label{eq:Z-toy-exact}
\ee
where $\B(x,y)=\Gamma(x)\Gamma(y)/\Gamma(x+y)$ is the Euler beta function.
This spectral zeta function admits a semi-classical expansion\footnote{%
The WKB expansion \eqref{eq:Z-toy-WKB} is slightly different from \eqref{eq:WKB} for ABJM spectral zeta.
The expansion \eqref{eq:Z-toy-WKB} is rather similar to the so-called $N_f$ matrix model in $N_f \to 0$ 
\cite{Mezei:2013gqa, GM, HO}.
} around $k=0$
\be
Z^\text{toy}(s)=\frac{2}{\pi^2s^2 k}-\frac{k}{48}+\frac{\zeta(3) s}{16\pi^2}k^2-\frac{\pi^2 s^2}{5120}k^3+\cO(k^4).
\label{eq:Z-toy-WKB}
\ee
Note that $Z^\text{toy}(s)$ does not receive any non-perturbative corrections in $k$.
We want to understand the large $\mu$ behavior of the grand potential in this toy model.
First, it is easy to see that $Z^\text{toy}(s)$ has infinite poles at
\be
s=-\frac{4m}{k},\qquad m=0,1,2,\dots.
\ee
There are no singularities except for these points on the whole complex plane.
Using the Mellin-Barnes representation \eqref{eq:J-MB}, one can know the large $\mu$ behavior of the grand potential.
In addition to the above poles, the integrand of the Mellin-Barnes integral has poles at $s=-n$ ($n=1,2,\dots$)
in the region $\real s<c$.
The pole at $s=0$ determines the leading behavior in the large $\mu$ limit.
It is easy to find
\be
\cJ^\text{toy}(\mu,k)=\frac{\mu^3}{3\pi^2 k}+\( \frac{1}{3k}-\frac{k}{48} \)\mu+\frac{\zeta(3)}{16\pi^2}k^2+
\cJ^\text{toy}_\text{M2}(\mu,k)+\cJ^\text{toy}_\text{WS}(\mu,k),
\label{eq:J-toy-large}
\ee
where $\cJ^\text{toy}_\text{M2}(\mu,k)$ and $\cJ^\text{toy}_\text{WS}(\mu,k)$ are subleading exponentially suppressed corrections in $\mu \to \infty$.
We separate these two contributions because their origins are slightly different.
The ``membrane instanton'' correction $\cJ^\text{toy}_\text{M2}(\mu,k)$ comes from the poles at $s=-n$.
The terminology is just an  analogy with the ABJM Fermi-gas.
We find that it is generically given by
\be
\cJ^\text{toy}_\text{M2}(\mu,k)= \sum_{n=1}^\infty \frac{(-1)^n}{(2\pi n)^2} \B\( -\frac{n k}{4}, -\frac{n k}{4} \) \re^{-n \mu}.
\label{eq:J-toy-M2}
\ee
On the other hand, the ``worldsheet instanton'' correction $\cJ^\text{toy}_\text{WS}(\mu,k)$ comes from the
non-perturbative poles $s=-4m/k$.
After a simple computation, one finds
\be
\cJ^\text{toy}_\text{WS}(\mu,k)=-\frac{k}{8\pi} \sum_{m=1}^\infty \frac{1}{m^2} \binom{2m}{m} \csc \( \frac{4m\pi}{k} \) \re^{-\frac{4m\mu}{k}}.
\label{eq:J-toy-WS}
\ee
The result \eqref{eq:J-toy-large} with \eqref{eq:J-toy-M2} and \eqref{eq:J-toy-WS} is the complete large $\mu$ expansion of the grand potential for the toy model.
In this example, we can completely control all the corrections.

Let us remark some of important lessons that we should learn from this toy model.

\paragraph{Pole cancellation.}
We notice that both $\cJ^\text{toy}_\text{M2}(\mu,k)$ and $\cJ^\text{toy}_\text{WS}(\mu,k)$ diverge at rational values of $k$.
As an example, let us consider the limit $k \to 4$.
In this limit, the membrane instanton part is given by
\be
\lim_{k \to 4} \cJ^\text{toy}_\text{M2}(\mu,k) =\sum_{n=0}^\infty \biggl[
-\frac{2(-1)^n}{n^3 \pi^2 (k-4)} \binom{2n}{n}+\frac{(-1)^n}{n^2 \pi^2}\binom{2n}{n} (H_{n}-H_{2n})+\cO(k-4) 
\biggr]\re^{-n \mu},
\ee
where $H_n$ is the $n$-th harmonic number.
The first term is divergent.
Similarly, the worldsheet instanton part is given by
\be
\lim_{k \to 4} \cJ^\text{toy}_\text{WS}(\mu,k) =\sum_{n=0}^\infty \biggl[
\frac{2(-1)^n}{n^3 \pi^2 (k-4)} \binom{2n}{n}+\frac{(-1)^n}{n^2 \pi^2}\binom{2n}{n} \(\frac{\mu}{2}+\frac{1}{n} \)+\cO(k-4) 
\biggr]\re^{-n \mu}.
\ee
The important point is that the sum of these two corrections is \textit{finite}
\be
\lim_{k \to 4} (\cJ^\text{toy}_\text{M2}+\cJ^\text{toy}_\text{WS} )
=\sum_{n=0}^\infty \frac{(-1)^n}{n^2 \pi^2} \binom{2n}{n} \( \frac{\mu}{2}+\frac{1}{n}+H_n-H_{2n} \) \re^{-n \mu}.
\label{eq:M2+WS-k=4}
\ee
This is, of course, not accidental.
At the level of the spectral zeta function or the Mellin-Barnes representation, 
there is no problem in the limit $k \to 4$.
It gives finite values for any $k$.
What happens in $k \to 4$? 
In this limit, the non-perturbative poles $s=-4n/k$ ``collide'' with the perturbative poles $s=-n$.
At $k=4$, both poles degenerate, and the orders of the poles become higher.
One can easily check that the spectral zeta function at $k=4$,
\be
Z_{k=4}^\text{toy}(s)=\frac{1}{4\pi^2 s} \B\(s,s\),
\ee
correctly reproduces the mixed corrections \eqref{eq:M2+WS-k=4}.

\paragraph{Non-perturbative effects.}
The toy model here is already involving.
At the level of the spectral zeta function, there are no non-perturbative corrections in $k$.
Everything can be understood only in the perturbative analysis.
Nevertheless, the grand potential receives the non-perturbative correction $\cJ^\text{toy}_\text{WS}(\mu,k)$ 
in the large $\mu$ regime.
The source of this correction is the non-perturbative poles at $s=-4m/k$ ($m=1,2,\dots$).
We stress that these poles are never visible in the semi-classical expansion around $k=0$.
In fact, using the WKB expansion \eqref{eq:Z-toy-WKB}, one encounters, at each order, only the perturbative poles of
the integrand in the Mellin-Barnes representation.
The non-perturbative poles appear after \textit{resumming all the perturbative corrections to} $Z^\text{toy}(s)$. 
In the next section, we will give a strong evidence that the similar mechanism 
also works in the ABJM Fermi-gas:
The perturbative resummation induces non-perturbative poles.

\section{Spectral zeta function in the ABJM Fermi-gas}\label{sec:ABJM-zeta}
Let us proceed to the ABJM Fermi-gas.
Unfortunately, unlike the toy model in the previous section, the exact spectral zeta function in this model is not known.%
\footnote{We should note that, at $k=1,2,4$, the exact spectrum can be computed by solving an 
\textit{exact quantization condition} \cite{KM, CGM, Kallen, Wang:2014ega, GHM2}. It would be interesting to analyze the exact spectral zeta on the complex plane for these special cases.}
Therefore we cannot apply the method in the previous section immediately.
The situation is very limited.
To explore its analytic property, we first consider the semi-classical expansion of $Z(s)$,
as in \eqref{eq:WKB}.
For $s=\ell$, ($\ell=1,2,\dots$), there is a powerful way to compute the semi-classical expansion of $Z(\ell)$
by using the TBA equations \cite{CM}.
Here, however, we take another strategy.
We use the Wigner transform in quantum mechanics.
This method was already considered in the original paper on the ABJM Fermi-gas \cite{MP2},
but we develop a bit more efficient way.
This approach is useful in the case where the TBA description is not known.
We next extrapolate the semi-classical expansion to the finite-$k$ regime by
using the Pad\'e approximation, and observe the pole structure of $Z(s)$.

\subsection{Semi-classical analysis: Wigner transform}
For a quantum mechanical operator $\hat{A}$, the Wigner transform is defined by
\be
A_\text{W}(x,p):=\int_{-\infty}^\infty \rd x' \re^{\frac{\ri px'}{\hbar}} \bra{x-\frac{x'}{2}} \hat{A} \ket{x+\frac{x'}{2}},
\ee
where recall that $\hbar=2\pi k$.
By definition, the diagonal element of $\hat{A}$ in the coordinate representation is given by
\be
\bra{x} \hat{A} \ket{x}=\int_{-\infty}^\infty \frac{\rd p}{2\pi \hbar} A_\text{W}(x,p).
\ee
Thus the trace is given by
\be
\Tr \hat{A}=\int_{-\infty}^\infty \rd x \bra{x}\hat{A} \ket{x}=\int_{-\infty}^\infty \frac{\rd x \rd p}{2\pi \hbar} A_\text{W}(x,p).
\ee
We first apply the Wigner transform to the inverse of density operator
\be
\hat{\cO}:=\hat{\rho}^{-1}=\(2\cosh \frac{\hat{x}}{2}\)^{1/2}\( 2\cosh \frac{\hat{p}}{2} \) \(2\cosh \frac{\hat{x}}{2}\)^{1/2}.
\ee
As derived in appendix~\ref{sec:Wigner}, the Wigner transform of $\hat{\cO}$ is \textit{exactly} given by
\be
\cO_\text{W}(x,p)=4\cosh \frac{p}{2} \( \cosh^2 \frac{x}{2}-\sin^2 \frac{\pi k}{4} \)^{1/2}.
\label{eq:O_W}
\ee
We want to compute
\be
Z(s)=\Tr \hat{\rho}^s=\Tr \re^{-s \hat{H}}=\Tr \hat{\cO}^{-s}.
\ee
We use a general argument in \cite{MP2}.
For a given function $f$, the Wigner transform of $f(\hat{\cO})$ is given by
\be
f(\hat{\cO})_\text{W}=\sum_{r=0}^\infty \frac{1}{r!}f^{(r)}(\cO_\text{W}) \cG_r,
\label{eq:f-W}
\ee
where
\be
\cG_r:=[(\hat{\cO}-\cO_\text{W})^r]_\text{W}.
\ee
In computing this quantity, one encounters the Wigner transform
of operator products.
The Wigner transform of a product of two operators is computed by
\be
(\hat{A}\cdot \hat{B})_\text{W}=A_\text{W}\star B_\text{W},
\ee
where the Moyal product $\star$ is defined by
\be
\ba
A \star B&:= A(x,p) \exp \left[ \frac{\ri \hbar}{2} ( \stackrel{\leftarrow}{\pd}_x \stackrel{\rightarrow}{\pd}_p
-\stackrel{\leftarrow}{\pd}_p\stackrel{\rightarrow}{\pd}_x  ) \right] B(x,p) \\
&=\sum_{n=0}^\infty \sum_{m=0}^n (-1)^m \binom{n}{m} \frac{1}{n!}\(\frac{\ri \hbar}{2}\)^n
\pd_x^m \pd_p^{n-m} A(x,p) \pd_p^m \pd_x^{n-m} B(x,p) .
\ea
\ee
In our purpose, we set $f(x)=x^{-s}$.
In this case, the Wigner transform \eqref{eq:f-W} becomes
\be
(\hat{\cO}^{-s})_\text{W}=\sum_{r=0}^\infty \frac{(-1)^r (s)_r}{r!} \cO_\text{W}^{-s-r} \cG_r,
\label{eq:f-W-2}
\ee
where $(s)_r$ is the Pochhammer symbol.
Using the exact relation \eqref{eq:O_W}, one can compute the WKB expansion
of this formula up to any order, in principle.
Then after the integration over $(x,p)$, one obtains $Z(s)$.
Note that the integral converges only for $\real s>0$.
After the integration, one can do the analytic continuation.

It is complicated to compute the integral of \eqref{eq:f-W-2} for general $s$.
Practically, it is sufficient to compute it for $s \in \mathbb{Z}_{>0}$.
From the lower order information (see \eqref{eq:Z-WKB}),
one can guess a form of higher order correction with finite unknown parameters.
These parameters are fixed only by the result for finite $s \in \mathbb{Z}_{>0}$.
If the ansatz is correct, the obtained result must reproduce the corrections for other values of $s$.
In this way, one can fix each coefficient in \eqref{eq:Z-WKB} order by order.
Equivalently, it is convenient to use an interesting observation in \cite{MN2}.
Recall that the WKB expansion of the grand potential is given by \eqref{eq:WKB}.
The quantum correction $\cJ^{(n)} (\mu)$ can be constructed by acting a non-trivial differential operator to
the classical one:
\be
\cJ^{(n)}  (\mu)=\cD^{(n)}  \cJ^{(0)} (\mu), \qquad n=1,2,\dots,
\label{eq:Jn}
\ee 
where $\cD^{(n)} $ is a differential operator of $\mu$.
Its explicit form up to $n=2$ is found in \cite{MN2}.
Using this observation, we have indeed fixed the differential operator up to $n=10$.
The result up to $n=4$ is given in appendix~\ref{sec:diff}.
The obtained grand potential in this way is in precise agreement with the one from the TBA approach \cite{CM}.\footnote{
We thank Kazumi Okuyama and Masazumi Honda for sharing their results on the higher order differential operators in another project \cite{HHO}.
}

Now let us consider the WKB expansion of the spectral zeta function, given by \eqref{eq:WKB}.
As already computed in section~\ref{sec:review}, 
the classical part is explicitly given by \eqref{eq:Z0}.
Combining \eqref{eq:J} and \eqref{eq:Jn}, one easily obtains the relation
\be
Z^{(0)} (\ell)\cD^{(n)}  (\re^{\ell \mu})=Z^{(n)} (\ell) \re^{\ell \mu},\qquad \ell=1,2,\dots.
\ee
This means that the correction $Z^{(n)} (\ell)$ is obtained by replacing $\pd_\mu$ in $\cD^{(n)}$ by $\ell$. 
It is very natural that the correction $Z^{(n)} (s)$ is analytically continued by
\be
Z^{(0)} (s)\cD^{(n)}  (\re^{s \mu})=Z^{(n)} (s) \re^{s \mu},\qquad s \in \mathbb{C}.
\ee 
Using the explicit form of $\cD^{(n)}$ in appendix~\ref{sec:diff}, the WKB expansion of $Z(s)$ is given by
\be
\ba
&Z_\text{WKB}(s)=\frac{Z^{(0)} (s)}{k} \biggl[ 1+\frac{ s^2(1-s)}{96 (1+s)}(\pi k)^2-\frac{ s^3(1-s) \left(16+27 s+7 s^2\right)}{92160 (1+s) (3+s)}(\pi k)^4 \\
&\quad+\frac{s^3 (1-s) \left(256+2560 s+3216 s^2+1649 s^3+376 s^4+31 s^5\right)}{61931520 (1+s) (3+s) (5+s)}(\pi k)^6+\cO(k^8)
\biggr].
\ea
\label{eq:Z-WKB}
\ee
We use this expansion in order to explore the pole structure of $Z(s)$ in the next subsection.

\subsection{Pole structure and non-perturbative effects}

In this subsection, we explore the pole structure of $Z(s)$ for finite $k$.
Since we do not know the all-order coefficients of the WKB expansion,
we cannot extrapolate it to the finite-$k$ regime, directly.
Here we apply the Pad\'e approximation to the WKB expansion \eqref{eq:Z-WKB}.
An advantage of the Pad\'e approximation is that it captures the pole structure of
the original function.
We see that the extrapolation of the Pad\'e approximant of \eqref{eq:Z-WKB} to finite $k$ strongly implies that $Z(s)$ has
a pole at $s=-4/k$, which is of course invisible as long as one considers 
only each term in \eqref{eq:Z-WKB}.

Let us first see that the spectral zeta function $Z(s)$ does not have any poles in $\real s >0$.
In this region, we can use the definition \eqref{eq:spec-zeta}.
We use an inequality
\be
|Z(s)|=\left| \sum_{n=0}^\infty \re^{-s E_n} \right| \leq \sum_{n=0}^\infty |\re^{-s E_n}|=\sum_{n=0}^\infty \re^{-x E_n},
\ee
where $x:=\real s$ and $E_n=-\log \lambda_n$ is the eigenvalues of the quantum Hamiltonian $\hat{H}$.
Since $E_n >0$ for $\forall n \in \mathbb{Z}_{\geq 0}$, the rightmost equation monotonically decreases w.r.t. $x$. 
We conclude that the rightmost equation is finite for $x>0$, and thus $Z(s)$ has no poles in $x>0$.

To see the poles in $\real s\leq 0$, let us consider the Pad\'e approximation of a given function
\be
f(x)=\sum_{n=0}^\infty f_n x^n.
\ee
We denote its Pad\'e approximant with numerator order $M$ and denominator order $N$ by
\be
[M/N]_f(x)=\frac{a_0+a_1 x+\cdots a_M x^M}{1+b_1 x+\cdots b_N x^N}.
\ee
The $M+N+1$ coefficients $\{ a_0,\dots,a_M;b_1,\dots,b_N \}$ are uniquely fixed by requiring 
\be
f(x)-[M/N]_f(x)=\cO(x^{M+N+1}).
\ee
The Pad\'e approximant is the most general form in the ratonal-type approximation.
It, of course, includes the original Taylor expansion.
It is well-known that the Pad\'e approximant captures the global structure of the original function,
compared to the Taylor expansion.

\paragraph{Test in the toy model.}

\begin{figure}[tb]
\begin{center}
\begin{tabular}{cc}
%\hspace{-3mm}
\resizebox{65mm}{!}{\includegraphics{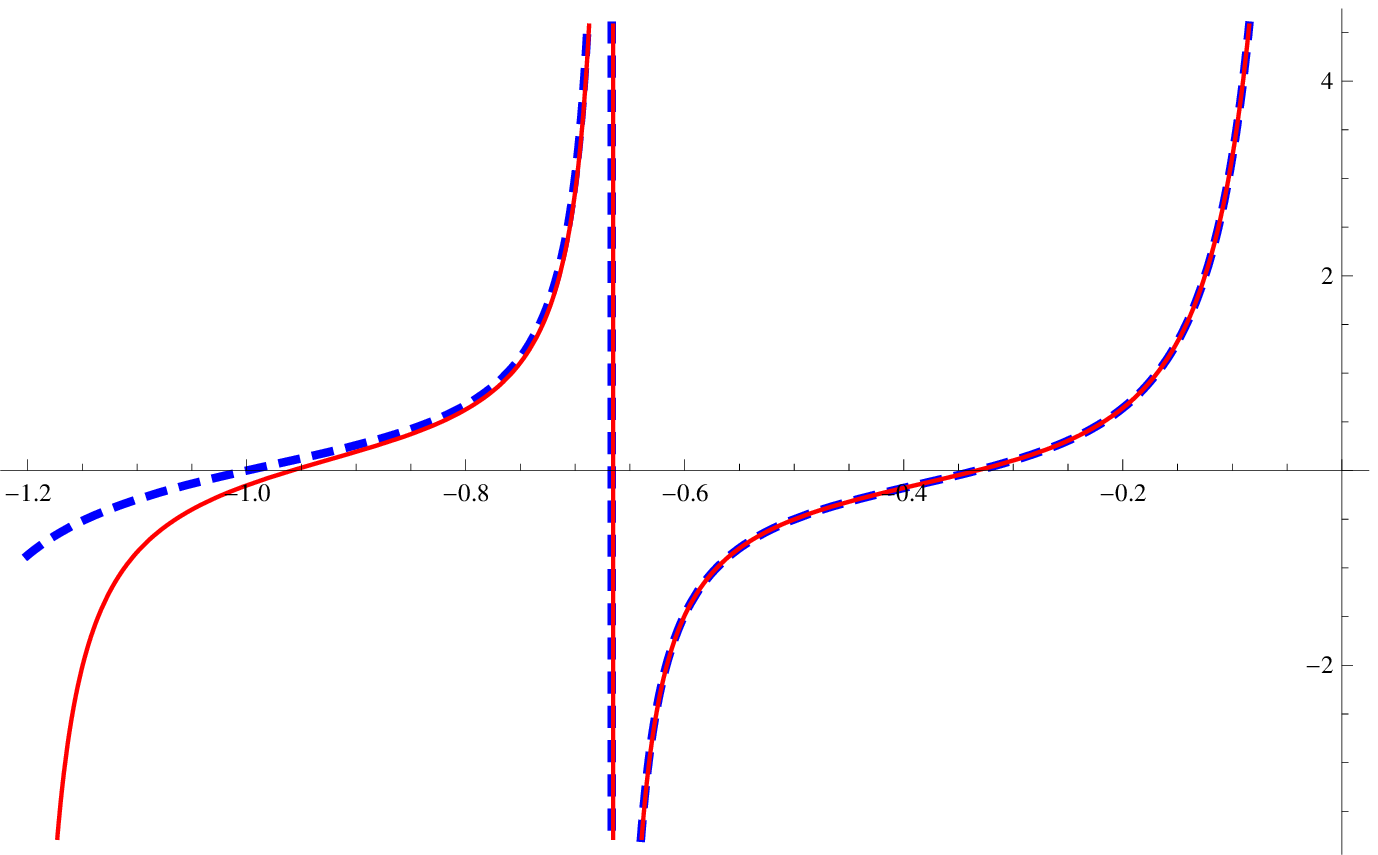}}
\hspace{4mm}
&
\resizebox{65mm}{!}{\includegraphics{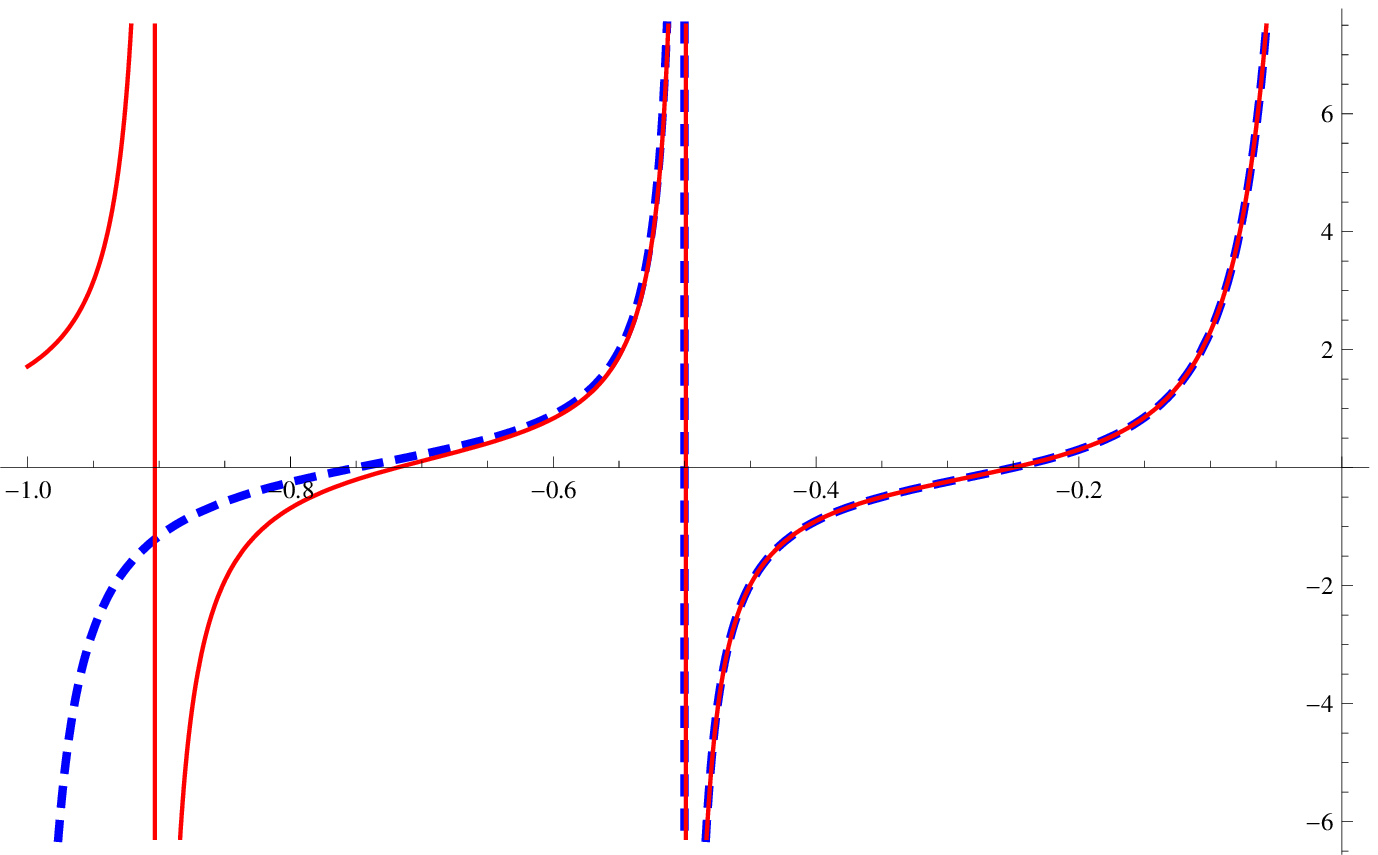}}
\vspace{-5mm}
\end{tabular}
\end{center}
  \caption{This figure shows the Pad\'e approximant $Z^{\text{toy},[5/5]}_\text{Pad\'e}(s;k)$ (Red solid line) 
and the exact function $Z^\text{toy}(s;k)$ (Blue dashed line) at $k=6$ (Left) and $k=8$ (Right) as functions of $s$.
The Pad\'e approximant captures a global structure of the original function (at least to the nearest pole from the origin).}
  \label{fig:Z-toy-Pade}
\end{figure}

Before proceeding to the ABJM spectral zeta, let us test a validity of this method
by using the toy model in the previous section.
We start with the WKB expansion \eqref{eq:Z-toy-WKB}.
For convenience, we denote this expansion by
\be
\ba
Z^\text{toy}(s;k)&=\frac{2}{\pi^2 s^2 k} \widehat{Z}_\text{WKB}^\text{toy}(s;k), \\
\widehat{Z}_\text{WKB}^\text{toy}(s;k)&=1-\frac{\pi^2 s^2 k^2}{96}+\frac{\zeta(3)s^3k^3}{32}
-\frac{\pi^4 s^4 k^4}{10240}+\cO(k^5).
\ea
\ee
Then, we apply the Pad\'e approximation to $\widehat{Z}_\text{WKB}^\text{toy}(s;k)$.
We also introduce a notation
\be
Z^{\text{toy}, [M/N]}_\text{Pad\'e}(s;k):= \frac{2}{\pi^2 s^2 k} [M/N]_{\widehat{Z}_\text{WKB}^\text{toy}}(s;k).
\ee
What we show here is to compare this approximant with the exact function \eqref{eq:Z-toy-exact} for finite $k$.
In figure~\ref{fig:Z-toy-Pade}, we show the behavior $Z^{\text{toy}, [5/5]}_\text{Pad\'e}(s;k)$ for $k=6,8$.
The red solid lines are the approximant $Z^{\text{toy}, [5/5]}_\text{Pad\'e}(s;k)$, while
the blue dashed ones are the exact  $Z^\text{toy}(s;k)$.  
Both graphs show good agreements in the regime $-4/k<s<0$.

\begin{figure}[tb]
\begin{center}
\begin{tabular}{cc}
%\hspace{-3mm}
\resizebox{65mm}{!}{\includegraphics{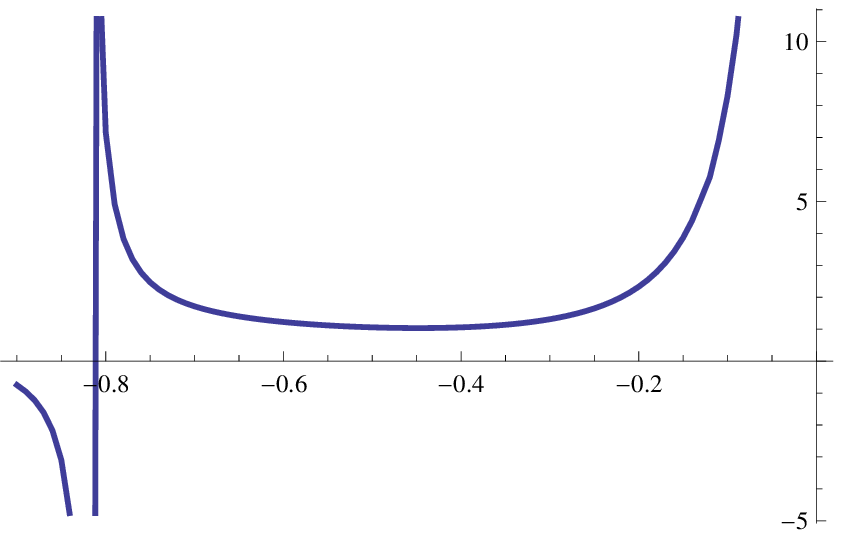}}
\hspace{4mm}
&
\resizebox{65mm}{!}{\includegraphics{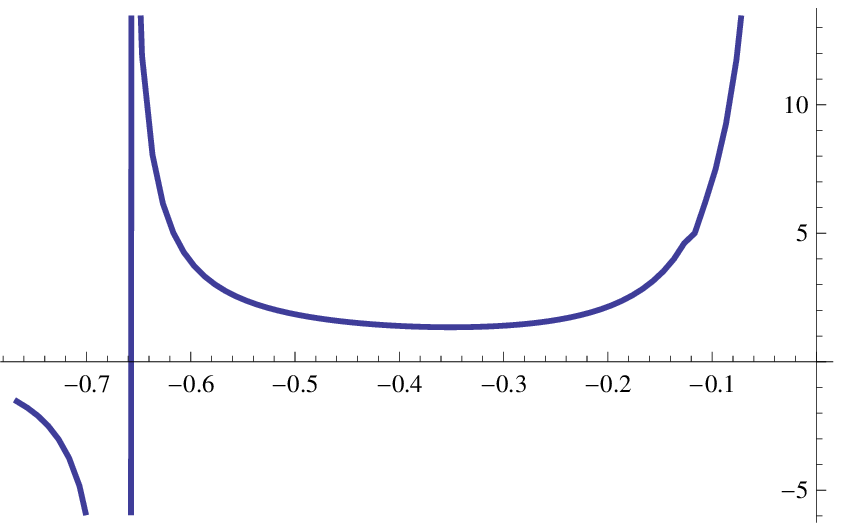}}
\vspace{-5mm}
\end{tabular}
\end{center}
  \caption{We show the Pade approximant $Z^{[10/10]}_\text{Pad\'e}(s;k)$ at $k=5$ (Left) and $k=6$ (Right) as a function of $s$. 
It is observed that they diverge at $s \approx -4/k$.}
  \label{fig:Z-Pade}
\end{figure}

\paragraph{Application to ABJM.}
Now let us apply the Pad\'e approximation to the spectral zeta function in the ABJM case.
As in the toy model, we first define
\be
\widehat{Z}_\text{WKB}(s;k):= \frac{k Z_\text{WKB}(s)}{Z^{(0)} (s)}=1+\frac{ s^2(1-s)}{96 (1+s)}(\pi k)^2+\cdots.
\label{eq:Z-WKB-norm}
\ee
We use the WKB expansion up to $\cO(k^{20})$ and compute the Pad\'e approximant $[10/10]_{\widehat{Z}_\text{WKB}}$.
We also denote
\be
Z^{[M/N]}_\text{Pad\'e}(s;k)=\frac{Z^{(0)}(s)}{k} [M/N]_{\widehat{Z}_\text{WKB}}(s;k).
\ee
In figure~\ref{fig:Z-Pade}, we show the behavior of $Z^{[10/10]}_\text{Pad\'e}(s;k)$ for $k=5,6$.
These graphs strongly suggest that $Z(s)$ has a pole at $s=-4/k$.
Using the Mellin-Barnes representation, this pole leads to the non-perturbative correction of order $\re^{-\frac{4\mu}{k}}$.
The result is completely consistent with the leading worldsheet instanton correction.

\paragraph{Non-perturbative corrections.}
It is clear that the coefficients of the non-perturbative corrections are related to the residues of the non-perturbative
poles.
If the spectral zeta function behaves near the non-perturbative pole at $s=-4m/k$ ($m=1,2,\dots$) as
\be
\lim_{s \to -4m/k} Z(s)=\frac{z_{m}^{(-1)}(k)}{s+4m/k}+\cO(1),
\ee
then the Mellin-Barnes representation states that the grand potential receives the correction
\be
\frac{\pi k}{4m}z_{m}^{(-1)}(k) \csc \( \frac{4m \pi}{k} \) \re^{-\frac{4m \mu}{k}},
\ee
where we have assumed that this pole do not overlap with any other poles.
To reproduce the known result\footnote{%
Recall that to compare the known result one should be careful about the contribution from the generalized theta function \cite{CGM, GHM1, GHM2}.
At this order, however, there is no such a contribution, and one can compare the result with the topological string one.
}
 in \cite{HMO2} the first residue $z_1^{(-1)}(k)$, for example, should take the form
\be
z_{1}^{(-1)}(k)=\frac{8}{\pi k} \cot \( \frac{2\pi}{k} \).
\ee
So far, we have no principle to determine such residues analytically.
It would be very interesting to understand the analyticity of $Z(s)$ in more detail.

\subsection{A comment on non-perturbative correction to spectral zeta function}
In the previous subsection, we saw that the WKB perturbative resummation produces the non-perturbative pole
that causes the leading non-perturbative correction to the grand potential.
We must note that there is a possibility that the spectral zeta function itself receives non-perturbative corrections 
of form $\re^{-1/k}$.
This type of correction is invisible in the semi-classical analysis. 
Let us give a comment on this possibility.
We first note that at $s=1,2$ the spectral zeta function is exactly computed in \cite{Okuyama}:
\be
Z(1)=\frac{1}{4k},\qquad
Z(2)=\frac{1}{4\pi^2}\int_0^\infty \rd t \frac{t}{\cosh^2 t \sinh kt}.
\ee
It is easy to see that these results do not receive the non-perturbative correction in the small $k$ limit.
Also, if plugging $s=-2,-4$ into \eqref{eq:Z-WKB-norm} (and using the result in appendix~\ref{sec:diff}), 
we get the following expansions:
\be
\ba
\widehat{Z}_\text{WKB}(s=-2;k)&=1-\frac{(\pi k)^2}{8}+\frac{(\pi k)^4}{384}-\frac{(\pi k)^6}{46080}
+\frac{(\pi k)^8}{10321920}+\cO(k^{10}),\\
\widehat{Z}_\text{WKB}(s=-4;k)&=1-\frac{5(\pi k)^2}{18}+\frac{5 (\pi k)^2}{216}-\frac{(\pi k)^6}{1296}
+\frac{(\pi k)^8}{72576}+\cO(k^{10}).
\ea
\ee
It is easy to guess that these are reproduced by
\be
\ba
\widehat{Z}_\text{WKB}(s=-2;k)&=\cos \frac{\pi k}{2}, \\
\widehat{Z}_\text{WKB}(s=-4;k)&=\frac{1}{9} (4+5 \cos \pi k).
\ea
\ee
Hence for $s=-2,-4$, the WKB sums can be performed exactly.\footnote{%
This is not accidental. In fact, the function $\widehat{Z}_\text{WKB}(s=-2\ell;k)$ are related to the function $a_\ell(k)$ 
in \eqref{eq:J-WKB-large}.
This is a consequence of the property \eqref{eq:Jn}.
We also note that at $s=-2\ell$, the classical part $Z^{(0)}(s)$ is divergent but $\widehat{Z}_\text{WKB}(s;k)$ is finite.
}
Obviously these do not have the non-perturbative corrections in $k$.
All of these results seem to imply that the spectral zeta function does not receive the non-perturbative corrections.
However, we cannot conclude it at present.
We need a more detailed analysis.
If the spectral zeta function has the non-perturbative correction, 
these corrections might also have the non-perturbative poles in \eqref{eq:poles}, in principle.
We hope to report a resolution of this issue in a future.

\section{Another example: topological string on the resolved conifold}\label{sec:top}
In this section, we give another example, for which our procedure works.
In particular, we will see that the topological string free energy on the resolved conifold is precisely reproduced
by our prescription in  a non-perturbative manner. 
This example is closely related to a quantum spectral problem in the topological string theory \cite{GHM1, Huang:2014eha}.
The setup here is explained in \cite{GHM1} in detail.

\subsection{General framework}
In \cite{GHM1}, a new perspective to formulate the topological strings was proposed, based on an earlier work \cite{ACDKV}
and the Fermi-gas formulation \cite{MP2}.
The key idea is to associate an operator of a quantized mirror curve with a spectral problem.
Our starting point is a mirror curve describing a toric Calabi-Yau threefold.
Here we restrict ourselves to the following simple case:
\be
W(\re^x,\re^p)=\cO_{m,n}(x,p)+\tilde{u}=0,
\label{eq:mirror-curve}
\ee
where $\cO_{m,n}(x,p)$ is defined by
\be
\cO_{m,n}(x,p):=\re^x+\re^p+\re^{-mx-np}.
\ee
As explained in \cite{Kashaev:2015kha, MZ}, this mirror curve describes the anti-canonical bundle of the weighted projective space
$\mathbb{P}(1,m,n)$.
Now we want to ``quantize'' this curve.
Following the prescription in \cite{ACDKV}, the quantized mirror curve is given by
\be
(\re^{\hat{x}}+\re^{\hat{p}}+\re^{-m \hat{x}-n\hat{p}}+\tilde{u})\ket{\Psi}=0,
\label{eq:q-mirror-curve}
\ee
where $\hat{x}$ and $\hat{p}$ satisfy the canonical commutation relation \eqref{eq:comm-rel}.
The wavefunction $\Psi(x)=\bra{x}\Psi \rangle$ describes a brane that probes the geometry \cite{ACDKV}.
As conjectured in \cite{GHM1} and proved in \cite{Kashaev:2015kha}, the inverse operator
\be
\hat{\rho}_{m,n}:=\hat{\cO}^{-1}_{m,n}=(\re^{\hat{x}}+\re^{\hat{p}}+\re^{-m \hat{x}-n\hat{p}})^{-1}, \qquad (m, n>0),
\ee 
is positive-definite and of trace class.
This means that the operator $\hat{\rho}_{m,n}$ has positive and discrete eigenvalues, as in the ABJM Fermi-gas.
In \cite{GHM1}, an explicit formula for a spectral determinant of this operator was conjectured (for some special cases).
In \cite{Kashaev:2015kha}, it was shown that this operator has a good representation in certain coordinate.
This representation is useful to reformulate the topological strings as matrix models \cite{MZ}.
Here we focus on the spectral zeta function $Z_{m,n}(s)$ of the operator $\hat{\rho}_{m,n}$.

As in sections~\ref{sec:review} and \ref{sec:ABJM-zeta}, we consider the semi-classical limit $\hbar \to 0$.
As before, the spectral zeta function and the grand potential have the following WKB expansion:
\be
Z_{m,n}^\text{WKB}(s)=\frac{1}{\hbar} \sum_{\ell=0}^\infty \hbar^{2\ell} Z_{m,n}^{(\ell)}(s),\qquad
\cJ_{m,n}^\text{WKB}(\mu,\hbar)=\frac{1}{\hbar} \sum_{\ell=0}^\infty \hbar^{2\ell} \cJ_{m,n}^{(\ell)}(\mu).
\ee
At the classical level, the spectral zeta function is computed by
\be
Z_{m,n}^{(0)}(s)=\int \!\frac{\rd x \rd p}{2\pi} \frac{1}{(\re^x+\re^p+\re^{-mx-np})^s}
=\frac{\Gamma(\frac{s}{m+n+1})\Gamma(\frac{ms}{m+n+1})\Gamma(\frac{ns}{m+n+1})}{2\pi (m+n+1)\Gamma(s)}.
\ee
Therefore the classical grand potential is given by
\be
\cJ_{m,n}^{(0)}(\mu)=-\frac{1}{2\pi} \int_{c-\ri \infty}^{c+\ri \infty} \frac{\rd s}{2\pi \ri} 
\Gamma(-s)\Gamma\(\frac{s}{m+n+1}\)\Gamma\(\frac{ms}{m+n+1}\)\Gamma\(\frac{ns}{m+n+1}\)\re^{s\mu}. 
\ee
In the large $\mu$ limit, one finds
\be
\ba
\cJ_{m,n}^{(0)}(\mu)&=\frac{(m+n+1)^2}{12\pi mn}\mu^3+\frac{\pi(m^2+mn+n^2+m+n+1)}{12mn}\mu \\
&\qquad+\frac{(m+1)(n+1)(m+n)\zeta(3)}{2\pi mn(m+n+1)}+\cO(\re^{-\frac{m+n+1}{m}\mu},\re^{-\frac{m+n+1}{n}\mu},\re^{-(m+n+1)\mu}),
\ea
\ee
where exponentially suppressed corrections come from the residues of the poles in $s<0$.

Next we consider the Wigner transform to compute the quantum corrections.
As in the previous section, the Wigner transform of the operator $\hat{\cO}_{m,n}$ can be computed exactly,
and it takes the very simple form
\be
(\hat{\cO}_{m,n})_\text{W}(x,p)=\re^{x}+\re^{p}+\re^{-mx-np}.
\ee
The Wigner transform is just the same form as the original function!
One can confirm this result by a direct computation along the line in appendix~\ref{sec:Wigner}.
Then using \eqref{eq:f-W-2}, one can compute the quantum corrections to $Z_{m,n}(s)$ systematically.
As in \eqref{eq:Jn}, it is useful to fix the differential operator acting on the grand potential.
In appendix~\ref{sec:diff}, we present the explicit forms of the first three corrections.
Using these operators, one finds the WKB expansion of $Z_{m,n}(s)$:
\be
\ba
Z_{m,n}^\text{WKB}(s)&=\frac{Z_{m,n}^{(0)}(s)}{\hbar}\biggl[ 
1-\frac{m n s^2}{24(m+n+1)}\hbar ^2\\
&\quad+\biggl(\frac{m n(m+1)(n+1)(m+n)s^3}{2880(m+n+1)^2}
+\frac{7 m^2n^2s^4}{5760(m+n+1)^2}\biggr)\hbar ^4+\cO(\hbar^6) \biggr],
\ea
\label{eq:Zmn-WKB}
\ee
and also the grand potential in the large $\mu$ limit:
\be
\cJ_{m,n}^{\text{WKB}}(\mu,\hbar)=\frac{C_{m,n}(\hbar)}{3}\mu^3+B_{m,n}(\hbar) \mu+A_{m,n}(\hbar)+\cdots,
%\cO(\re^{-\frac{m+n+1}{m}\mu},\re^{-\frac{m+n+1}{n}\mu},\re^{-(m+n+1)\mu})
\ee
where $\cdots$ denotes the exponentially suppressed corrections and
\be
\ba
C_{m,n}(\hbar)&=\frac{(m+n+1)^2}{4\pi mn \hbar}, \\
B_{m,n}(\hbar)&=\frac{\pi(m^2+mn+n^2+m+n+1)}{12mn \hbar}-\frac{(m+n+1)\hbar}{48\pi}.
\ea
\ee
Finally we conjecture the form of $A_{m,n}(\hbar)$ with the full quantum corrections:
\be
A_{m,n}(\hbar)=\frac{1}{4} \left[ A_\text{c}\( \frac{\hbar}{\pi} \) +A_\text{c}\( \frac{m\hbar}{\pi} \)+A_\text{c}\( \frac{n\hbar}{\pi} \)
-A_\text{c}\( \frac{(m+n+1)\hbar}{\pi}\) \right] ,
\ee
where $A_\text{c}(k)$ is just the same function appearing in \eqref{eq:A-int}.
Using the differential operators in appendix~\ref{sec:diff}, this conjecture is checked up to $\hbar^5$.
It is not easy to determine exact forms of the exponentially suppressed corrections.
All of these results are valid for any $m>0$ and $n>0$, for which the spectral problem for $\hat{\rho}_{m,n}$ is well-defined.
In particular, for $m=n=1$, the corresponding geometry is local $\mathbb{P}^2$.
In this case, all the results here reproduce the ones in \cite{GHM1}.
In the next subsection, we analytically continue them to $n<0$.
In this case, the spectral problem is no longer well-defined, i.e., the operator $\hat{\rho}_{m,n}$ does not have
the discrete spectrum.
Nevertheless, this continuation provides us a very interesting result.
In particular, we can see the appearance of the non-perturbative poles analytically.

\subsection{Analytic continuation}
Looking at the differential operators \eqref{eq:Dop-top}, 
one notices that they drastically simplify if setting $n=-1$ (or $m=-1$) or $n=-m$.
Let us first consider the case of $n=-1$. The computation for $n=-m$ is almost same.
In the case of $n=-1$, the differential operators in \eqref{eq:Dop-top} reduce to
\be
\cD_{m,-1}^{(1)}=\frac{1}{24}\pd_\mu^2,\qquad
\cD_{m,-1}^{(2)}=\frac{7}{5760}\pd_\mu^4,\qquad
\cD_{m,-1}^{(3)}=\frac{31}{967680} \pd_\mu^6.
\ee
The result does not depend on $m$.
Quite interestingly, the same operators also appear in the quantum mechanical system with inverted harmonic potential 
in the computation of the quantum periods in the $c=1$ string \cite{ACDKV}.
Currently, we do not have a clear understanding of this agreement because the setups look quite different.
The WKB expansion \eqref{eq:Zmn-WKB} also simplifies:
\be
Z_{m,-1}^\text{WKB}(s)=\frac{Z_{m,-1}^{(0)}(s)}{\hbar}\biggl( 
1+\frac{s^2}{24}\hbar^2+\frac{7s^4}{5760}\hbar^4+\frac{31s^6}{967680}\hbar^6+\cO(\hbar^8) \biggr),
\label{eq:Zmm1-WKB-1}
\ee
where the classical part is given by
\be
Z_{m,-1}^{(0)}(s)=-\frac{1}{2s}\csc \( \frac{\pi s}{m} \).
\ee
It is an easy guess work to find the all-order result to reproduce \eqref{eq:Zmm1-WKB-1}.
The result is the following
\be
Z_{m,-1}^\text{WKB}(s)=\frac{Z_{m,-1}^{(0)}(s)}{\hbar}\cdot \frac{\hbar s}{2} \csc \( \frac{\hbar s}{2} \)
=-\frac{1}{4}\csc \( \frac{\pi s}{m} \) \csc\( \frac{\hbar s}{2} \).
\label{eq:Zmm1}
\ee
We have to note that this zeta function has poles in $\real s>0$ unlike the ABJM spectral zeta.
This is of course because of the analytic continuation to $n<0$.
As mentioned previously, for $n<0$, the spectral problem is no longer well-defined.
Thus we cannot apply the argument in the previous section to show that the spectral zeta has no poles in $\real s>0$.
However, as we will see below, this zeta function gives a remarkable result.
In the following, we forget a physical interpretation in $\mu \to -\infty$, and concentrate on the large $\mu$ behavior.
It is easy to see that \eqref{eq:Zmm1} has non-perturbative poles at $s=2\pi \ell/\hbar$ ($\ell \in \mathbb{Z}$).
We stress that these poles appear after the all-order resummation of the WKB expansion.

We formally plug the result \eqref{eq:Zmm1} into the Mellin-Barnes integral \eqref{eq:J-MB}:
\be
\cJ_{m,-1}(\mu,\hbar)=-\int_{c-\ri \infty}^{c+\ri \infty} \frac{\rd s}{2\pi \ri} 
\frac{\pi}{4s} \csc(\pi s) \csc \( \frac{\pi s}{m} \) \csc \( \frac{\hbar s}{2} \) \re^{s\mu}.
\ee
Now it is clear that the integrand has three-type of poles: two kinds of perturbative poles at $s=0,\pm 1, \pm 2,\dots$ and
at $s=\pm m, \pm 2m, \dots$ and non-perturbative poles at $s=\pm 2\pi/\hbar, \pm 4\pi/\hbar,\dots$.
In the large $\mu$ limit, we obtain
\be
\cJ_{m,-1}=-\frac{m}{12\pi \hbar} \mu^3-\( \frac{\pi (m+m^{-1})}{12\hbar}+\frac{m \hbar}{48\pi} \)\mu
+\cJ_{m,-1}^\text{M2,I}+\cJ_{m,-1}^\text{M2,II}+\cJ_{m,-1}^\text{WS},
\ee
where $\cJ_{m,-1}^\text{M2,I}$ and $\cJ_{m,-1}^\text{M2,II}$ are the contributions from the two kinds of the perturbative poles:
\be
\ba
\cJ_{m,-1}^\text{M2,I}&=-\sum_{\ell=1}^\infty \Res_{s=-\ell} \frac{\pi}{4s} \csc(\pi s) \csc \( \frac{\pi s}{m} \) \csc \( \frac{\hbar s}{2} \) \re^{s\mu} \\
&=\sum_{\ell=1}^\infty \frac{(-1)^\ell}{4\ell} \csc \( \frac{ \pi \ell}{m} \) \csc \( \frac{\ell \hbar}{2} \) \re^{-\ell \mu},\\
\cJ_{m,-1}^\text{M2,II}&=-\sum_{\ell=1}^\infty \Res_{s=-\ell m} \frac{\pi}{4s} \csc(\pi s) \csc \( \frac{\pi s}{m} \) \csc \( \frac{\hbar s}{2} \) \re^{s\mu} \\
&=\sum_{\ell=1}^\infty \frac{(-1)^\ell}{4\ell} \csc ( \pi \ell m)\csc \( \frac{\ell m \hbar}{2} \) \re^{-\ell m \mu}.
\ea
\ee
Also, $\cJ_{m,-1}^\text{WS}$ is the contribution from the non-perturbative poles:
\be
\ba
\cJ_{m,-1}^\text{WS}&=-\sum_{\ell=1}^\infty \Res_{s=-2\pi\ell/\hbar} \frac{\pi}{4s} \csc(\pi s) \csc \( \frac{\pi s}{m} \) \csc \( \frac{\hbar s}{2} \) \re^{s\mu} \\
&=\sum_{\ell=1}^\infty \frac{(-1)^\ell}{4\ell} \csc \( \frac{ 2\pi^2 \ell}{\hbar} \) \csc \( \frac{2\pi^2 \ell}{m\hbar} \) \re^{-\frac{2\pi \ell}{\hbar} \mu}.
\ea
\ee
These are the complete large $\mu$ expansions in the case of $n=-1$.

Next, let us proceed to the case of $n=-m$.
In this case, the differential operators reduce to
\be
\cD_{m,-m}^{(1)}=\frac{m^2}{24}\pd_\mu^2,\qquad
\cD_{m,-m}^{(2)}=\frac{7m^4}{5760} \pd_\mu^4,\qquad
\cD_{m,-m}^{(3)}=\frac{31m^6}{967680}\pd_\mu^6.
\ee
The conjectural all-order spectral zeta function is thus given by
\be
Z_{m,-m}^\text{WKB}(s)=-\frac{1}{4}\csc( \pi m s) \csc \( \frac{m \hbar s}{2} \).
\ee
Repeating the same computation above, one finally gets
\be
\cJ_{m,-m}=-\frac{ \mu^3}{12\pi m^2 \hbar}-\( \frac{\pi(1+m^{-2})}{12\hbar}+\frac{\hbar}{48\pi} \)\mu
+\cJ_{m,-m}^\text{M2,I}+\cJ_{m,-m}^\text{M2,II}+\cJ_{m,-m}^\text{WS},
\ee
where
\be
\ba
\cJ_{m,-m}^\text{M2,I}&=\sum_{\ell=1}^\infty \frac{(-1)^\ell}{4\ell} \csc ( \pi \ell m ) \csc \( \frac{\ell m \hbar}{2} \) \re^{-\ell \mu},\\
\cJ_{m,-m}^\text{M2,II}&=\sum_{\ell=1}^\infty \frac{(-1)^\ell}{4\ell} \csc \( \frac{\pi \ell}{m}\) \csc \( \frac{\ell \hbar}{2} \) \re^{-\frac{\ell}{m} \mu},\\
\cJ_{m,-m}^\text{WS}&=\sum_{\ell=1}^\infty \frac{(-1)^\ell}{4\ell} \csc \( \frac{ 2\pi^2 \ell}{\hbar} \) \csc \( \frac{2\pi^2 \ell}{m\hbar} \) 
\re^{-\frac{2\pi \ell}{m\hbar} \mu}.
\ea
\ee

\paragraph{Resolved conifold limit.}
To make contact with the topological string result in the literature,
we further take the limit $m \to 1$.
In this limit, one should be careful about the membrane instanton corrections since both $\cJ^\text{M2,I}$ and $\cJ^\text{M2,II}$
diverge in $m \to 1$.
As seen in section~\ref{sec:toy}, such divergences are, however, canceled by each other.
The sum of these two contributions is finite.
After a simple computation, one obtains
\be
\ba
\cJ_{1,-1}^\text{M2}(\mu, \hbar)&=\lim_{m \to 1} (\cJ_{m,-1}^\text{M2,I}+\cJ_{m,-1}^\text{M2,II})
=\lim_{m \to 1} (\cJ_{m,-m}^\text{M2,I}+\cJ_{m,-m}^\text{M2,II}) \\
&=\sum_{\ell=1}^\infty (\widetilde{b}_\ell(\hbar) \mu+\widetilde{c}_\ell(\hbar) ) \re^{-\ell \mu},
\label{eq:J-M2-coni}
\ea
\ee
where
\be
\ba
\widetilde{b}_\ell(\hbar)&=-\frac{1}{4\pi \ell} \csc \( \frac{\ell \hbar}{2} \),\\
\widetilde{c}_\ell(\hbar)&=-\frac{1}{4\pi \ell^2} \csc \( \frac{\ell \hbar}{2} \)
\left[ \frac{\ell \hbar}{2} \cot \( \frac{\ell \hbar}{2} \)+1 \right].
\ea
\ee
There is no limit problem for the worldsheet instanton correction:
\be
\cJ_{1,-1}^\text{WS}(\mu, \hbar)=\sum_{\ell=1}^\infty \frac{(-1)^\ell}{4\ell} \csc^2 \( \frac{ 2\pi^2 \ell}{\hbar} \)
\re^{-\frac{2\pi \ell}{\hbar} \mu}.
\label{eq:J-WS-coni}
\ee
Let us return to the geometrical meaning of this limit.
It is known that the mirror curve \eqref{eq:mirror-curve} with $(m,n)=(1,-1)$ describes the resolved conifold 
(see \cite{HKRS} for example).
Therefore these results should be compared with the topological string free energy on the resolved conifold.
In the next subsection, we will see that the results \eqref{eq:J-M2-coni} and \eqref{eq:J-WS-coni} are indeed reproduced by the refined
topological string on the resolved conifold, following the procedure in \cite{HMMO}. 
This test is a strong evidence of our proposal that the perturbative resummation of $Z(s)$
captures not only the perturbative corrections to $J(\mu)$ but also the non-perturbative corrections in $\hbar$.

\subsection{Comparison with the topological string on the resolved conifold}
In this subsection, we compare the results \eqref{eq:J-M2-coni} and \eqref{eq:J-WS-coni} with the
topological string on the resolved conifold.
We start with the free energy of the refined topological string on the resolved conifold with two parameters $(\epsilon_1,\epsilon_2)$,
\be
F(\epsilon_1,\epsilon_2;Q)=-\sum_{\ell=1}^\infty \frac{Q^\ell}{\ell(q^{\ell/2}-q^{-\ell/2})(t^{\ell/2}-t^{-\ell/2})},
\ee
where
\be
q=\re^{\epsilon_1},\qquad t=\re^{-\epsilon_2}.
\ee
We first see that the worldsheet instanton correction \eqref{eq:J-WS-coni} is reproduced from the  unrefined topological string
free energy, i.e.,
\be
 \epsilon_1=-\epsilon_2 \qquad \text{i.e.} \qquad
q=t=q_s,\qquad Q=Q_s.
\ee
In this slice, the free energy is given by
\be
F_\text{top}(q_s;Q_s)=-\sum_{\ell=1}^\infty \frac{Q_s^\ell}{\ell(q_s^{\ell/2}-q_s^{-\ell/2})^2}.
\ee
Following \cite{HMMO}, we identify the parameters
\be
q_s=\re^{2\pi \ri \lambda}%=\re^{\frac{4\pi^2 \ri}{\hbar}}
,\qquad
Q_s=\re^{T+\pi \ri}%=-\re^{-\frac{2\pi}{\hbar} \mu}.
,
\label{eq:para3}
\ee
where the string coupling $\lambda$ and the K\"ahler modulus $T$ are related to $\hbar$ and $\mu$ by
\be
\lambda=\frac{2\pi}{\hbar},\qquad
T=- \lambda \mu=-\frac{2\pi \mu}{\hbar}.
\label{eq:para2}
\ee
Then the free energy is finally given by
\be
F_\text{top}(q_s;Q_s)=\sum_{\ell=1}^\infty \frac{(-1)^\ell}{4\ell} \csc^2 \( \frac{2\pi^2 \ell}{\hbar} \) \re^{-\frac{2\pi \ell}{\hbar} \mu}.
\label{eq:F-top}
\ee
This is in perfect agreement with \eqref{eq:J-WS-coni}.

Next we consider the so-called Nekrasov-Shatashvili (NS) limit \cite{NS}: $\epsilon_2 \to 0$.
In this limit, the free energy reduces to
\be
F_\text{NS}(q;Q):=\lim_{\epsilon_2 \to 0} \epsilon_2 F(\epsilon_1,\epsilon_2;Q)
=\sum_{\ell=1}^\infty \frac{1}{\ell^2} \frac{Q^\ell}{q^{\ell/2}-q^{-\ell/2}}.
\ee
Following \cite{HMMO} again, we identify the parameters as follows:
\be
q=\re^{\frac{2\pi \ri}{\lambda}}=\re^{\ri \hbar}, \qquad Q=\re^{\frac{T}{\lambda}}=\re^{-\mu},
\label{eq:para1}
\ee
where $\lambda$ and $T$ is the same parameters above.
Then the free energy is written as
\be
F_\text{NS}\(\frac{1}{\lambda},\frac{T}{\lambda}\)=\frac{1}{2\ri} \sum_{\ell=1}^\infty 
\frac{\re^{\frac{\ell T}{\lambda}}}{\ell^2 \sin (\frac{\pi \ell}{\lambda} )}.
\ee
Now let us apply the procedure in \cite{HMMO}.
This procedure claims that the membrane instanton correction to the free energy is constructed from the free
energy in the NS limit,
\be
\ba
F_\text{M2}&=\frac{1}{2\pi \ri} \frac{\pd}{\pd \lambda} \left[ \lambda F_\text{NS}\(\frac{1}{\lambda},\frac{T}{\lambda}\) \right]\\
&=\sum_{\ell=1}^\infty \frac{1}{4\pi \ell^2} \csc\( \frac{\pi \ell}{\lambda} \)
\left[ \frac{\ell T}{\lambda}-\frac{\pi \ell}{\lambda} \cot \( \frac{\pi \ell}{\lambda} \)-1 \right]\re^{\frac{\ell T}{\lambda}}.
\ea
\ee
Therefore one finally gets
\be
F_\text{M2}=-\sum_{\ell=1}^\infty \frac{1}{4\pi \ell^2} \csc \( \frac{\ell \hbar}{2} \)
\left[ \ell \mu+\frac{\ell \hbar}{2} \cot\( \frac{\ell \hbar}{2} \)+1 \right] \re^{-\ell \mu}.
\label{eq:F-M2}
\ee
This also agrees with \eqref{eq:J-M2-coni}, perfectly.
Obviously, these agreements are highly non-trivial.

Let us explain a significance of these results.
In \cite{HMMO}, a non-perturbative completion of the topological string free energy was proposed.
This proposal claims that the sum
\be
F_\text{top}(q_s;Q_s)+F_\text{M2}(q;Q)
\label{eq:F-full}
\ee
should compute the complete unrefined topological string free energy, including all the non-perturbative corrections.
On one hand, the standard large $N$ limit of the unrefined topological string corresponds to $\lambda \to 0$ or $\hbar \to \infty$
with $T$ or $\mu/\hbar$ held fixed in \eqref{eq:para2}.
In this limit, the genus expansion or its Gopakumar-Vafa resummation captures the first term in \eqref{eq:F-full}.
The second term contributes as a non-perturbative correction of order $\re^{-1/\lambda}$ (see \eqref{eq:para1}).
In particular, the Gopakumar-Vafa resummation \eqref{eq:F-top} has an infinite number of poles on the real axis of $\hbar$.
These poles are precisely canceled by the membrane instanton correction \eqref{eq:F-M2} as shown for general backgrounds in \cite{HMMO}.
As a consequence, the sum \eqref{eq:F-full} is always well-defined for any $\hbar$ or $\lambda$.
On the other hand, starting with the quantized mirror curve \eqref{eq:q-mirror-curve}, a natural parameter is of course
the Planck constant $\hbar$.
The semi-classical limit is $\hbar \to 0$.
In this picture, the second term in \eqref{eq:F-full} is naturally computed.
The remarkable fact is that this quantum system is just described by the refined topological strings in the NS limit \cite{ACDKV, NS}.
The first term appears as 
a quantum mechanical non-perturbative correction \cite{KM}.
Therefore the unrefined topological strings are closely related to the refined topological strings in the NS limit
via an ``S-duality''.
We are seeing the same quantity \eqref{eq:F-full} from different perspectives.%
\footnote{A similar but more symmetric situation is also found in vortex-antivortex factorization \cite{Pasquetti}.
In the perturbative regime of the vortex partition function, the antivortex appears as a non-perturbative correction and vice versa.
Here, the roles of vortex and antivortex are played by the worldsheet instantons and the membrane instantons.
}
This is a main consequence of \cite{HMMO, KM}.

Let us recall that in the computation in the previous subsection we considered only the perturbative resummation $Z_{m,n}^\text{WKB}(s)$.
The non-perturbative corrections to $J_{m,n}(\mu)$ naturally appear as the non-perturbative poles in the integrand of \eqref{eq:J-MB}
after the resummation.
The fact that the results \eqref{eq:J-M2-coni} and \eqref{eq:J-WS-coni} completely reproduces \eqref{eq:F-full}, 
proposed in \cite{HMMO}, strongly
implies that the perturbative resummation of $Z(s)$ captures all the non-perturbative information on $J(\mu)$, precisely.

\paragraph{Remark.} 
The non-perturbative effects in the topological string on the resolved conifold have already been studied 
in a different approach \cite{PS}.
The analysis in \cite{PS} is based on the Borel analysis and closely related to the resurgence theory \cite{Ecalle}.
See \cite{Dorigoni}, for instance, for a pedagogical introduction to the resurgence.
The Borel analysis shows that the perturbative genus expansion of the resolved conifold 
free energy is an asymptotic series and non-Borel summable. 
It has a non-perturbative ambiguity\footnote{%
In a spirit of the resurgence, this ambiguity must be canceled by the other contributions in
the complete trans-series expansion.
However, the ambiguity (or the discontinuity of the lateral Borel resummation) itself has important non-perturbative information.
} 
of the Borel resummation
because of the Stokes phenomenon.
In \cite{PS}, the discontinuity of the free energy across the Stokes line was computed.
The discontinuity is given by
\be
\text{Disc}\, F_\text{conifold}=-\frac{\ri}{2\pi g_s}\sum_{\ell=1}^\infty \sum_{m \in \mathbb{Z}} 
\( \frac{2\pi(t+2\pi \ri m)}{\ell}+\frac{g_s}{\ell^2} \)
\re^{-\frac{2\pi \ell(t+2\pi \ri m)}{g_s}},
\label{eq:disc-1}
\ee
where $g_s$ is the string coupling and $t$ is the K\"ahler modulus.
This result is understood as a Schwinger effect and thus considered to compute the BPS pair-production rate.
Here we compare this result with the membrane instanton correction $F_\text{M2}(q;Q)$.\footnote{%
We thank Marcos Mari\~no for suggesting this interesting comparison.
}
To do so, we first have to identify the parameters.
To match the perturbative free energy \eqref{eq:F-top} with the one in \cite{PS},
we have to set
\be
g_s=2\pi \lambda=\frac{4\pi^2}{\hbar},\qquad t=-T-\pi \ri=\frac{2\pi \mu}{\hbar}-\pi \ri. 
\ee
Next, we split the sum over $m$ in \eqref{eq:disc-1} into two parts:
\be
\ba
\text{Disc}\, F_\text{conifold}&=-\frac{\ri}{2\pi \lambda}\sum_{\ell=1}^\infty \sum_{m=1}^\infty
\( \frac{-T+(2m-1)\pi \ri}{\ell}+\frac{\lambda}{\ell^2} \) \re^{-\frac{\ell}{\lambda}(-T+(2m-1)\pi \ri)}\\
&\quad-\frac{\ri}{2\pi \lambda}\sum_{\ell=1}^\infty \sum_{m=1}^\infty
\( \frac{-T-(2m-1)\pi \ri}{\ell}+\frac{\lambda}{\ell^2} \) \re^{-\frac{\ell}{\lambda}(-T-(2m-1)\pi \ri)}.
\ea
\label{eq:disc-2}
\ee
After perfoming the sum over $m$, the first term is evaluated as
\be
\ba
&-\frac{\ri}{2\pi \lambda}\sum_{\ell=1}^\infty \sum_{m=1}^\infty
\( \frac{-T+(2m-1)\pi \ri}{\ell}+\frac{\lambda}{\ell^2} \) \re^{-\frac{\ell}{\lambda}(-T+(2m-1)\pi \ri)}\\
&\quad=-\sum_{\ell=1}^\infty \frac{1}{4\pi \ell^2} \csc \( \frac{\ell \hbar}{2} \)
\left[ \ell \mu+\frac{\ell \hbar}{2} \cot\( \frac{\ell \hbar}{2} \)+1 \right] \re^{-\ell \mu}.
\ea 
\ee
This perfectly coincides with the membrane instanton correction \eqref{eq:F-M2}!
Similarly, the second term is the same, but with opposite sign, contribution.
Note that in deriving \eqref{eq:disc-1}, the refined topological string in the NS limit is of course not used at all.
It is remarkable that these two fairly independent computations precisely agree.
However, naively the sum of the two contributions  in \eqref{eq:disc-2} totally vanishes. 
Currently we do not understand a definite reason of this cancellation.
It is very important to clarify it more clearly.

\section{Conclusions}\label{sec:conclusion}
In this paper, we proposed a new perspective on non-perturbative effects in the ABJM Fermi-gas system.
Our starting point is the spectral problem, and it naturally introduces the spectral zeta function.
The Mellin-Barnes type representation \eqref{eq:J-MB} describes the grand potential from the small $\kappa$ regime 
to the large $\kappa$ regime.
The spectral zeta function plays the essential role in this approach.
In particular, it is important to understand its pole structure.
The consistency condition with the known results requires that the spectral zeta function must have the non-perturbative
poles, as in \eqref{eq:poles}.
These poles cause the non-perturbative corrections to the grand potential.
So far, we do not have a systematic way to determine its pole structure, but in this paper,
we gave a strong evidence that it indeed has the non-perturbative pole at $s=-4/k$.
Quite interestingly, this non-perturbative pole appears after resumming the perturbative corrections to $Z(s)$.
Therefore we can naturally explain the appearance of the worldsheet instanton correction $\re^{-\frac{4\mu}{k}}$
in the grand potential from the perturbative analysis of $Z(s)$.
It would be significant to explore the pole structure in more detail.
It is also important to clarify whether $Z(s)$ itself receives the non-perturbative corrections or not.

We also gave another example on the topological string on the resolved conifold.
The perturbative resummation of the spectral zeta function perfectly reproduces the non-perturbative
proposal in \cite{HMMO} (and also \cite{PS}).
It would be interesting to push this analysis in more general cases.

The Mellin-Barnes representation can be widely applied to many other examples.
In fact, it also works in the topological string analysis formulated in \cite{GHM1, Kashaev:2015kha, MZ},
as seen for some special cases in this paper,
and in circular quiver Chern-Simons matrix models \cite{MN1, MN2, MN2, HHO} including \cite{Mezei:2013gqa, GM, HO}.
In these examples, it is hard to perform the sum \eqref{eq:J} even at the classical level.
The representation \eqref{eq:J-MB} allows us to compute the large $\mu$ expansion very systematically.
We emphasize that this representation should be valid at the quantum level.
It would be important to clarify the pole structure of the spectral zeta function in those models.

In the ABJM Fermi-gas, the consistency requires the pole structure \eqref{eq:poles}.
Naively, we expect that $Z(s)$ contains a factor
\be
\prod_{\ell, m=0}^\infty \frac{1}{s+2\ell+\frac{4m}{k}}.
\label{eq:double-gamma}
\ee
A natural function with this factor is the double gamma function $\Gamma_2(s|\omega_1,\omega_2)$
with $\omega_1=2$ and $\omega_2=4/k$.
Also, as in \eqref{eq:J-M2WS}, the grand potential is simplified by introducing the effective chemical potential \eqref{eq:mu-eff}.
This simplification seems to imply that the poles of $Z(s)$ are factorized into
\be
\prod_{\ell=1}^\infty \frac{1}{s+2\ell} \prod_{m=1}^\infty \frac{1}{s+\frac{4m}{k}},
\ee
rather than \eqref{eq:double-gamma}.
It would be nice to consider how such a simplification is understood in the Mellin-Barnes representation,
and also to investigate a relation between the spectral zeta function and the (double) gamma function
(or its cousins).
We hope to report all of these issues in a near future.

\acknowledgments{I thank D.~Dorigoni, M.~Honda, M.~Mari\~no and K.~Okuyama for valuable discussions and comments. 
I am especially grateful to M. Mari\~{n}o for reading the manuscript carefully and giving many helpful comments and suggestions.
}

\appendix

\section{Computing the Wigner transform}\label{sec:Wigner}
In this appendix, we derive \eqref{eq:O_W}.
By definition, the Wigner transform of $\hat{\cO}$ is given by
\be
\ba
\cO_\text{W}(x,p)=\int_{-\infty}^\infty \rd x' \re^{\frac{\ri px'}{\hbar}} \(2\cosh \frac{x+x'/2}{2}\)^{1/2}
\(2\cosh \frac{x-x'/2}{2}\)^{1/2} \bra{x-\frac{x'}{2}}2\cosh \frac{\hat{p}}{2} \ket{x+\frac{x'}{2}}.
\ea
\ee
The last part is written as
\be
 \bra{x-\frac{x'}{2}}2\cosh \frac{\hat{p}}{2} \ket{x+\frac{x'}{2}}=\int_{-\infty}^\infty \frac{\rd p'}{2\pi \hbar} \re^{-\frac{\ri p' x'}{\hbar}}
 2\cosh \frac{p'}{2}.
\ee
Therefore,
\be
\cO_\text{W}(x,p)=\int_{-\infty}^\infty \frac{\rd x' \rd p'}{(2\pi)^2} 
\re^{\frac{\ri (p-p')x'}{2\pi}}  2\(\cosh^2 \frac{x}{2}+\sinh^2 \frac{k x'}{4}\)^{1/2} 2\cosh \frac{p'}{2},
\ee
where we have rescaled the integration variable $x' \to k x'$.
Next, we expand the integrand around $k=0$.
We symbolically expand
\be
\(\cosh^2 \frac{x}{2}+\sinh^2 \frac{k x'}{4}\)^{1/2}=\sum_{m=0}^\infty c_m(x) (k x')^{2m}.
\ee
The important fact is that the integral over $x'$ gives the derivative of the delta function:
\be
\int_{-\infty}^\infty \frac{\rd x'}{2\pi} \re^{\frac{\ri (p-p')x'}{2\pi}} (x')^{n}=(-2\pi \ri)^{n} \delta^{(n)}{(p-p')}.
\ee
Then one can easily perform the integral over $p'$
\be
\ba
\int_{-\infty}^\infty \frac{\rd p'}{2\pi}(2\pi \ri)^{2m} \delta^{(2m)}{(p-p')}2\cosh \frac{p'}{2}
=(2\pi \ri \pd_p)^{2m} 2\cosh \frac{p}{2} 
=(\pi \ri )^{2m} 2\cosh \frac{p}{2}.
\ea
\ee
Using these results, we finally get
\be
\cO_\text{W}(x,p)=4\cosh \frac{p}{2} \sum_{m=0}^\infty c_m(x) (\pi \ri k)^{2m}
=4\cosh \frac{p}{2}\(\cosh^2 \frac{x}{2}-\sin^2 \frac{\pi k}{4}\)^{1/2}.
\ee
We note that this computation is exact, and $\cO_\text{W}(x,p)$ contains all the quantum corrections.

\section{Explicit results on differential operators}\label{sec:diff}
\subsection{ABJM Fermi-gas}
Here we list the forms of differential operators $\cD^{(n)}$ up to $n=4$:
\be
\ba
\cD^{(1)}&=\frac{\pi ^2(1-\pd_\mu) \pd_\mu^2 }{96 (1+\pd_\mu)},\\
\cD^{(2)}&=-\frac{ \pi ^4(1-\pd_\mu) \pd_\mu^3 \left(16+27 \pd_\mu+7 \pd_\mu^2\right)}{92160 (1+\pd_\mu) (3+\pd_\mu)},\\
\cD^{(3)}&=\frac{ \pi ^6(1-\pd_\mu) \pd_\mu^3 \left(256+2560 \pd_\mu+3216 \pd_\mu^2+1649 \pd_\mu^3+376 \pd_\mu^4
+31 \pd_\mu^5\right)}{61931520 (1+\pd_\mu) (3+\pd_\mu) (5+\pd_\mu)},\\
\cD^{(4)}&=-\frac{\pi^8(1-\pd_\mu) \pd_\mu^3}{39636172800 (1+\pd_\mu) (3+\pd_\mu) (5+\pd_\mu) (7+\pd_\mu)}
\bigl(20480+544768 \pd_\mu\\
&\qquad+1202688 \pd_\mu^2+1135872 \pd_\mu^3+584480 \pd_\mu^4+177609 \pd_\mu^5+31897 \pd_\mu^6\\
&\qquad+3119 \pd_\mu^7+127 \pd_\mu^8\bigr).
\ea
\ee
A basic strategy to fix these operators is as follows.
We first compute the expansion of $\cJ^{(n)}(\kappa)$ around $\kappa=0$.
This can be done by using the formula \eqref{eq:f-W-2} for $s=1,2,\dots$.
Taking an ansatz of the form of $\cD^{(n)}$, we try to fix unknown parameters to match the first
several coefficients of $\cJ^{(n)}(\kappa)$.
If the ansatz is correct, the obtained result must reproduce higher coefficients.
In this way, one can verify the obtained operator up to any desired order.

\subsection{Topological strings for three-term operator}
For the spectral problem considered in section~\ref{sec:top}, the differential operators are given by
\be
\ba
\cD_{m,n}^{(1)}&=-\frac{m n}{24(m+n+1)} \pd_\mu^2, \\
\cD_{m,n}^{(2)}&=\frac{m n(m+1)(n+1)(m+n)}{2880(m+n+1)^2}\pd_\mu^3+\frac{7 m^2n^2}{5760(m+n+1)^2}\pd_\mu^4, \\
\cD_{m,n}^{(3)}&=-\frac{m n(m+1)(n+1)(m+n)\left(m^2+m n+n^2+m+n+1\right)}{362880(m+n+1)^2}\pd_\mu^3 \\
&\quad-\frac{m n(m+1)^2(n+1)^2(m+n)^2}{120960(m+n+1)^3}\pd_\mu^4
-\frac{41m^2n^2(m+1)(n+1)(m+n)}{1451520(m+n+1)^3}\pd_\mu^5 \\
&\quad-\frac{31m^3n^3}{967680(m+n+1)^3}\pd_\mu^6.
\ea
\label{eq:Dop-top}
\ee
As noted in section~\ref{sec:top}, these operators drastically simplify in $m \to -1$ or $n \to -1$ or $m+n \to 0$.


\begin{thebibliography}{00}

 \bibitem{ABJM}
 O.~Aharony, O.~Bergman, D.~L.~Jafferis and J.~Maldacena, ``N=6 superconformal Chern-Simons-matter theories, M2-branes and their gravity duals,''
  JHEP {\bf 0810}, 091 (2008)
  [arXiv:0806.1218 [hep-th]].
  %%CITATION = JHEPA,0810,091;%%

\bibitem{Pestun} 
  V.~Pestun,
  ``Localization of gauge theory on a four-sphere and supersymmetric Wilson loops,''
  Commun.\ Math.\ Phys.\  {\bf 313}, 71 (2012)
  [arXiv:0712.2824 [hep-th]].
  %%CITATION = ARXIV:0712.2824;%%
  %481 citations counted in INSPIRE as of 15 Mar 2015

\bibitem{Teschner:2014oja} 
  J.~Teschner,
  ``Exact results on N=2 supersymmetric gauge theories,''
  arXiv:1412.7145 [hep-th].
  %%CITATION = ARXIV:1412.7145;%%
  %4 citations counted in INSPIRE as of 17 Mar 2015

  \bibitem{KWY1}
A.~Kapustin, B.~Willett and I.~Yaakov, ``Exact Results for Wilson Loops in Superconformal Chern-Simons Theories with Matter,''
  JHEP {\bf 1003}, 089 (2010)
  [arXiv:0909.4559 [hep-th]].
  %%CITATION = JHEPA,1003,089;%% 

\bibitem{Jafferis} 
  D.~L.~Jafferis,
  ``The Exact Superconformal R-Symmetry Extremizes Z,''
  JHEP {\bf 1205}, 159 (2012)
  [arXiv:1012.3210 [hep-th]].
  %%CITATION = ARXIV:1012.3210;%%
  %213 citations counted in INSPIRE as of 28 Jun 2014

\bibitem{HHL1} 
  N.~Hama, K.~Hosomichi and S.~Lee,
  ``Notes on SUSY Gauge Theories on Three-Sphere,''
  JHEP {\bf 1103}, 127 (2011)
  [arXiv:1012.3512 [hep-th]].
  %%CITATION = ARXIV:1012.3512;%%
  %158 citations counted in INSPIRE as of 28 Jun 2014

\bibitem{Hosomichi} 
  K.~Hosomichi,
  ``A review on SUSY gauge theories on $S^3$,''
  arXiv:1412.7128 [hep-th].
  %%CITATION = ARXIV:1412.7128;%%
  %2 citations counted in INSPIRE as of 15 Mar 2015
 
\bibitem{DMP1} 
  N.~Drukker, M.~Marino and P.~Putrov,
  ``From weak to strong coupling in ABJM theory,''
  Commun.\ Math.\ Phys.\  {\bf 306}, 511 (2011)
  [arXiv:1007.3837 [hep-th]].
  %%CITATION = ARXIV:1007.3837;%%
  %162 citations counted in INSPIRE as of 25 Jun 2014


\bibitem{MP1} 
  M.~Marino and P.~Putrov,
  ``Exact Results in ABJM Theory from Topological Strings,''
  JHEP {\bf 1006}, 011 (2010)
  [arXiv:0912.3074 [hep-th]].
  %%CITATION = ARXIV:0912.3074;%%
  %69 citations counted in INSPIRE as of 25 Jun 2014


\bibitem{GMZ} 
  A.~Grassi, M.~Marino and S.~Zakany,
  ``Resumming the string perturbation series,''
  JHEP {\bf 1505}, 038 (2015)
  [arXiv:1405.4214 [hep-th]].
  %%CITATION = ARXIV:1405.4214;%%
  %15 citations counted in INSPIRE as of 15 Oct 2015

\bibitem{DMP2} 
  N.~Drukker, M.~Marino and P.~Putrov,
  ``Nonperturbative aspects of ABJM theory,''
  JHEP {\bf 1111}, 141 (2011)
  [arXiv:1103.4844 [hep-th]].
  %%CITATION = ARXIV:1103.4844;%%
  %47 citations counted in INSPIRE as of 25 Jun 2014

\bibitem{Ecalle}
J.~Ecalle, Les Fonctions Resurgentes, vol. I-III. Publ. Math. Orsay, 1981.

\bibitem{Marino2008} 
  M.~Marino,
  ``Nonperturbative effects and nonperturbative definitions in matrix models and topological strings,''
  JHEP {\bf 0812}, 114 (2008)
  [arXiv:0805.3033 [hep-th]].
  %%CITATION = ARXIV:0805.3033;%%
  %58 citations counted in INSPIRE as of 25 mar 2015

\bibitem{Marino2006} 
  M.~Marino,
  ``Open string amplitudes and large order behavior in topological string theory,''
  JHEP {\bf 0803}, 060 (2008)
  [hep-th/0612127].
  %%CITATION = HEP-TH/0612127;%%
  %93 citations counted in INSPIRE as of 25 Mar 2015

\bibitem{MSW} 
  M.~Marino, R.~Schiappa and M.~Weiss,
  ``Nonperturbative Effects and the Large-Order Behavior of Matrix Models and Topological Strings,''
  Commun.\ Num.\ Theor.\ Phys.\  {\bf 2}, 349 (2008)
  [arXiv:0711.1954 [hep-th]].
  %%CITATION = ARXIV:0711.1954;%%
  %56 citations counted in INSPIRE as of 25 mar 2015

\bibitem{GIKM} 
  S.~Garoufalidis, A.~Its, A.~Kapaev and M.~Marino,
  ``Asymptotics of the instantons of Painleve I,''
  International Mathematics Research Notices (2012) Issue 3, pp.
  561-606
  [arXiv:1002.3634 [math.CA]].
  %%CITATION = ARXIV:1002.3634;%%
  %24 citations counted in INSPIRE as of 25 Mar 2015

\bibitem{ASV} 
  I.~Aniceto, R.~Schiappa and M.~Vonk,
  ``The Resurgence of Instantons in String Theory,''
  Commun.\ Num.\ Theor.\ Phys.\  {\bf 6}, 339 (2012)
  [arXiv:1106.5922 [hep-th]].
  %%CITATION = ARXIV:1106.5922;%%
  %28 citations counted in INSPIRE as of 15 Mar 2015


\bibitem{SV} 
  R.~Schiappa and R.~Vaz,
  ``The Resurgence of Instantons: Multi-Cut Stokes Phases and the Painleve II Equation,''
  Commun.\ Math.\ Phys.\  {\bf 330}, 655 (2014)
  [arXiv:1302.5138 [hep-th]].
  %%CITATION = ARXIV:1302.5138;%%
  %18 citations counted in INSPIRE as of 15 Mar 2015

\bibitem{SESV} 
  R.~C.~Santamaria, J.~D.~Edelstein, R.~Schiappa and M.~Vonk,
  ``Resurgent Transseries and the Holomorphic Anomaly,''
  arXiv:1308.1695 [hep-th].
  %%CITATION = ARXIV:1308.1695;%%
  %13 citations counted in INSPIRE as of 26 Mar 2015

\bibitem{ARS} 
I.~Aniceto, J.~G.~Russo and R.~Schiappa,
  ``Resurgent Analysis of Localizable Observables in Supersymmetric Gauge Theories,''
  JHEP {\bf 1503}, 172 (2015)
  [arXiv:1410.5834 [hep-th]].
  %%CITATION = ARXIV:1410.5834;%%
  %13 citations counted in INSPIRE as of 15 Oct 2015


\bibitem{HKPT} 
  C.~P.~Herzog, I.~R.~Klebanov, S.~S.~Pufu and T.~Tesileanu,
  ``Multi-Matrix Models and Tri-Sasaki Einstein Spaces,''
  Phys.\ Rev.\ D {\bf 83}, 046001 (2011)
  [arXiv:1011.5487 [hep-th]].
  %%CITATION = ARXIV:1011.5487;%%
  %92 citations counted in INSPIRE as of 28 Jun 2014

\bibitem{MP2} 
  M.~Marino and P.~Putrov,
  ``ABJM theory as a Fermi gas,''
  J.\ Stat.\ Mech.\  {\bf 1203}, P03001 (2012)
  [arXiv:1110.4066 [hep-th]].
  %%CITATION = ARXIV:1110.4066;%%
  %73 citations counted in INSPIRE as of 04 Feb 2015

\bibitem{HMO2} 
  Y.~Hatsuda, S.~Moriyama and K.~Okuyama,
  ``Instanton Effects in ABJM Theory from Fermi Gas Approach,''
  JHEP {\bf 1301}, 158 (2013)
  [arXiv:1211.1251 [hep-th]].
  %%CITATION = ARXIV:1211.1251;%%
  %18 citations counted in INSPIRE as of 25 Jun 2014


\bibitem{HO} 
  Y.~Hatsuda and K.~Okuyama,
  ``Probing non-perturbative effects in M-theory,''
  JHEP {\bf 1410}, 158 (2014)
  [arXiv:1407.3786 [hep-th]].
  %%CITATION = ARXIV:1407.3786;%%
  %10 citations counted in INSPIRE as of 15 mar 2015


\bibitem{MN1} 
  S.~Moriyama and T.~Nosaka,
  ``Partition Functions of Superconformal Chern-Simons Theories from Fermi Gas Approach,''
  JHEP {\bf 1411}, 164 (2014)
  [arXiv:1407.4268 [hep-th]].
  %%CITATION = ARXIV:1407.4268;%%
  %4 citations counted in INSPIRE as of 16 mar 2015

%\cite{Moriyama:2014waa}
\bibitem{MN2} 
  S.~Moriyama and T.~Nosaka,
  ``ABJM membrane instanton from a pole cancellation mechanism,''
  Phys.\ Rev.\ D {\bf 92}, no. 2, 026003 (2015)
  [arXiv:1410.4918 [hep-th]].
  %%CITATION = ARXIV:1410.4918;%%
  %7 citations counted in INSPIRE as of 15 Oct 2015


\bibitem{MN3} 
  S.~Moriyama and T.~Nosaka,
  ``Exact Instanton Expansion of Superconformal Chern-Simons Theories from Topological Strings,''
  JHEP {\bf 1505}, 022 (2015)
  [arXiv:1412.6243 [hep-th]].
  %%CITATION = ARXIV:1412.6243;%%
  %9 citations counted in INSPIRE as of 15 Oct 2015


\bibitem{HMMO} 
  Y.~Hatsuda, M.~Marino, S.~Moriyama and K.~Okuyama,
  ``Non-perturbative effects and the refined topological string,''
  JHEP {\bf 1409}, 168 (2014)
  [arXiv:1306.1734 [hep-th]].
  %%CITATION = ARXIV:1306.1734;%%
  %53 citations counted in INSPIRE as of 15 Oct 2015

\bibitem{KM} 
  J.~Kallen and M.~Marino,
  ``Instanton effects and quantum spectral curves,''
  arXiv:1308.6485 [hep-th].
  %%CITATION = ARXIV:1308.6485;%%
  %8 citations counted in INSPIRE as of 12 May 2014

\bibitem{GHM1} 
  A.~Grassi, Y.~Hatsuda and M.~Marino,
  ``Topological Strings from Quantum Mechanics,''
  arXiv:1410.3382 [hep-th].
  %%CITATION = ARXIV:1410.3382;%%
  %7 citations counted in INSPIRE as of 15 Mar 2015

\bibitem{ABJ} 
  O.~Aharony, O.~Bergman and D.~L.~Jafferis,
  ``Fractional M2-branes,''
  JHEP {\bf 0811}, 043 (2008)
  [arXiv:0807.4924 [hep-th]].
  %%CITATION = ARXIV:0807.4924;%%
  %302 citations counted in INSPIRE as of 05 Jul 2014


\bibitem{MaMo} 
  S.~Matsumoto and S.~Moriyama,
  ``ABJ Fractional Brane from ABJM Wilson Loop,''
  JHEP {\bf 1403}, 079 (2014)
  [arXiv:1310.8051 [hep-th]].
  %%CITATION = ARXIV:1310.8051;%%
  %15 citations counted in INSPIRE as of 16 mar 2015

\bibitem{HoO} 
  M.~Honda and K.~Okuyama,
  ``Exact results on ABJ theory and the refined topological string,''
  JHEP {\bf 1408}, 148 (2014)
  [arXiv:1405.3653 [hep-th]].
  %%CITATION = ARXIV:1405.3653;%%
  %14 citations counted in INSPIRE as of 16 Mar 2015

\bibitem{AHS} 
  H.~Awata, S.~Hirano and M.~Shigemori,
  ``The Partition Function of ABJ Theory,''
  Prog.\  Theor.\  Exp.\  Phys.\ , 053B04 (2013)
  [arXiv:1212.2966].
  %%CITATION = ARXIV:1212.2966;%%
  %22 citations counted in INSPIRE as of 15 Mar 2015

\bibitem{Honda} 
  M.~Honda,
  ``Direct derivation of "mirror" ABJ partition function,''
  JHEP {\bf 1312}, 046 (2013)
  [arXiv:1310.3126 [hep-th]].
  %%CITATION = ARXIV:1310.3126;%%
  %14 citations counted in INSPIRE as of 16 Mar 2015



\bibitem{Kashaev:2015kha} 
  R.~Kashaev and M.~Marino,
  ``Operators from mirror curves and the quantum dilogarithm,''
  arXiv:1501.01014 [hep-th].
  %%CITATION = ARXIV:1501.01014;%%
  %2 citations counted in INSPIRE as of 15 mar 2015

\bibitem{MZ} 
  M.~Marino and S.~Zakany,
  ``Matrix models from operators and topological strings,''
  arXiv:1502.02958 [hep-th].
  %%CITATION = ARXIV:1502.02958;%%

\bibitem{Drukker:2015awa} 
  N.~Drukker and J.~Felix,
  ``3d mirror symmetry as a canonical transformation,''
  JHEP {\bf 1505}, 004 (2015)
  [arXiv:1501.02268 [hep-th]].
  %%CITATION = ARXIV:1501.02268;%%
  %6 citations counted in INSPIRE as of 15 Oct 2015


\bibitem{HMO1} 
  Y.~Hatsuda, S.~Moriyama and K.~Okuyama,
  ``Exact Results on the ABJM Fermi Gas,''
  JHEP {\bf 1210}, 020 (2012)
  [arXiv:1207.4283 [hep-th]].
  %%CITATION = ARXIV:1207.4283;%%
  %20 citations counted in INSPIRE as of 25 Jun 2014


\bibitem{PY} 
  P.~Putrov and M.~Yamazaki,
  ``Exact ABJM Partition Function from TBA,''
  Mod.\ Phys.\ Lett.\ A {\bf 27}, 1250200 (2012)
  [arXiv:1207.5066 [hep-th]].
  %%CITATION = ARXIV:1207.5066;%%
  %16 citations counted in INSPIRE as of 25 Jun 2014

\bibitem{FHM} 
  H.~Fuji, S.~Hirano and S.~Moriyama,
  ``Summing Up All Genus Free Energy of ABJM Matrix Model,''
  JHEP {\bf 1108}, 001 (2011)
  [arXiv:1106.4631 [hep-th]].
  %%CITATION = ARXIV:1106.4631;%%
  %52 citations counted in INSPIRE as of 17 Mar 2015

\bibitem{KEK} 
  M.~Hanada, M.~Honda, Y.~Honma, J.~Nishimura, S.~Shiba and Y.~Yoshida,
  ``Numerical studies of the ABJM theory for arbitrary N at arbitrary coupling constant,''
  JHEP {\bf 1205}, 121 (2012)
  [arXiv:1202.5300 [hep-th]].
  %%CITATION = ARXIV:1202.5300;%%
  %33 citations counted in INSPIRE as of 25 Jun 2014


\bibitem{CM}
  F.~Calvo and M.~Mari\~no, ``Membrane instantons from a semiclassical TBA,''
  JHEP {\bf 1305}, 006 (2013)
  [arXiv:1212.5118 [hep-th]].
  %%CITATION = ARXIV:1212.5118;%% 

\bibitem{HMO3} 
  Y.~Hatsuda, S.~Moriyama and K.~Okuyama,
  ``Instanton Bound States in ABJM Theory,''
  JHEP {\bf 1305}, 054 (2013)
  [arXiv:1301.5184 [hep-th]].
  %%CITATION = ARXIV:1301.5184;%%
  %15 citations counted in INSPIRE as of 25 Jun 2014


\bibitem{CGM} 
  S.~Codesido, A.~Grassi and M.~Marino,
  ``Exact results in N=8 Chern-Simons-matter theories and quantum geometry,''
  arXiv:1409.1799 [hep-th].
  %%CITATION = ARXIV:1409.1799;%%
  %6 citations counted in INSPIRE as of 15 Mar 2015

\bibitem{GHM2} 
  A.~Grassi, Y.~Hatsuda and M.~Marino,
  ``Quantization conditions and functional equations in ABJ(M) theories,''
  arXiv:1410.7658 [hep-th].
  %%CITATION = ARXIV:1410.7658;%%
  %3 citations counted in INSPIRE as of 15 Mar 2015

\bibitem{CSSV} 
  R.~Couso-Santamaria, R.~Schiappa and R.~Vaz,
  ``Finite $N$ from Resurgent Large $N$,''
  Annals Phys.\  {\bf 356}, 1 (2015)
  [arXiv:1501.01007 [hep-th]].
  %%CITATION = ARXIV:1501.01007;%%

\bibitem{Voros}
  A.~Voros, ``Spectral zeta functions,'' Adv.~Stud.~Pure~Math~21. 327 (1992) 358.

\bibitem{ACDKV} 
  M.~Aganagic, M.~C.~N.~Cheng, R.~Dijkgraaf, D.~Krefl and C.~Vafa,
  ``Quantum Geometry of Refined Topological Strings,''
  JHEP {\bf 1211}, 019 (2012)
  [arXiv:1105.0630 [hep-th]].
  %%CITATION = ARXIV:1105.0630;%%
  %73 citations counted in INSPIRE as of 20 mar 2015

\bibitem{Huang:2014eha} 
  M.~x.~Huang and X.~f.~Wang,
  ``Topological Strings and Quantum Spectral Problems,''
  JHEP {\bf 1409}, 150 (2014)
  [arXiv:1406.6178 [hep-th]].
  %%CITATION = ARXIV:1406.6178;%%
  %11 citations counted in INSPIRE as of 20 Mar 2015


\bibitem{MarinoReview}
  M.~Mari\~no, ``Localization at large $N$ in Chern-Simons-matter theories,'' to appear.

\bibitem{HNS} 
  S.~Hirano, K.~Nii and M.~Shigemori,
  ``ABJ Wilson loops and Seiberg Duality,''
  arXiv:1406.4141 [hep-th].
  %%CITATION = ARXIV:1406.4141;%%
  %3 citations counted in INSPIRE as of 15 Mar 2015



\bibitem{Mezei:2013gqa} 
  M.~Mezei and S.~S.~Pufu,
  ``Three-sphere free energy for classical gauge groups,''
  JHEP {\bf 1402}, 037 (2014)
  [arXiv:1312.0920 [hep-th], arXiv:1312.0920].
  %%CITATION = ARXIV:1312.0920;%%
  %3 citations counted in INSPIRE as of 25 Jun 2014


\bibitem{GM} 
  A.~Grassi and M.~Marino,
  ``M-theoretic matrix models,''
  arXiv:1403.4276 [hep-th].
  %%CITATION = ARXIV:1403.4276;%%
  %1 citations counted in INSPIRE as of 23 Apr 2014


\bibitem{Kallen} 
  J.~Kallen,
  ``The spectral problem of the ABJ Fermi gas,''
  arXiv:1407.0625 [hep-th].
  %%CITATION = ARXIV:1407.0625;%%
  %10 citations counted in INSPIRE as of 17 mar 2015

\bibitem{Wang:2014ega} 
  X.~f.~Wang, X.~Wang and M.~x.~Huang,
  ``A Note on Instanton Effects in ABJM Theory,''
  JHEP {\bf 1411}, 100 (2014)
  [arXiv:1409.4967 [hep-th]].
  %%CITATION = ARXIV:1409.4967;%%
  %5 citations counted in INSPIRE as of 15 mar 2015
 

\bibitem{HHO}
  Y.~Hatsuda, M.~Honda and K.~Okuyama,
  ``Large N non-perturbative effects in $\mathcal{N}=4$ superconformal Chern-Simons theories,''
  JHEP {\bf 1509}, 046 (2015)
  [arXiv:1505.07120 [hep-th]].
  %%CITATION = ARXIV:1505.07120;%%
  %3 citations counted in INSPIRE as of 15 Oct 2015

\bibitem{Okuyama} 
  K.~Okuyama,
  ``A Note on the Partition Function of ABJM theory on $S^3$,''
  Prog.\ Theor.\ Phys.\  {\bf 127}, 229 (2012)
  [arXiv:1110.3555 [hep-th]].
  %%CITATION = ARXIV:1110.3555;%%
  %27 citations counted in INSPIRE as of 25 Jun 2014


\bibitem{HKRS} 
  M.~x.~Huang, A.~Klemm, J.~Reuter and M.~Schiereck,
  ``Quantum geometry of del Pezzo surfaces in the Nekrasov-Shatashvili limit,''
  JHEP {\bf 1502}, 031 (2015)
  [arXiv:1401.4723 [hep-th]].
  %%CITATION = ARXIV:1401.4723;%%
  %9 citations counted in INSPIRE as of 20 Mar 2015

\bibitem{NS} 
  N.~A.~Nekrasov and S.~L.~Shatashvili,
  ``Quantization of Integrable Systems and Four Dimensional Gauge Theories,''
  arXiv:0908.4052 [hep-th].
  %%CITATION = ARXIV:0908.4052;%%
  %234 citations counted in INSPIRE as of 20 Mar 2015

\bibitem{Pasquetti} 
  S.~Pasquetti,
  ``Factorisation of N = 2 Theories on the Squashed 3-Sphere,''
  JHEP {\bf 1204}, 120 (2012)
  [arXiv:1111.6905 [hep-th]].
  %%CITATION = ARXIV:1111.6905;%%
  %51 citations counted in INSPIRE as of 20 mar 2015

\bibitem{PS} 
  S.~Pasquetti and R.~Schiappa,
  ``Borel and Stokes Nonperturbative Phenomena in Topological String Theory and c=1 Matrix Models,''
  Annales Henri Poincare {\bf 11}, 351 (2010)
  [arXiv:0907.4082 [hep-th]].
  %%CITATION = ARXIV:0907.4082;%%
  %41 citations counted in INSPIRE as of 22 mar 2015

\bibitem{Dorigoni} 
  D.~Dorigoni,
  ``An Introduction to Resurgence, Trans-Series and Alien Calculus,''
  arXiv:1411.3585 [hep-th].
  %%CITATION = ARXIV:1411.3585;%%
  %5 citations counted in INSPIRE as of 22 mar 2015

\end{thebibliography}
\end{document}